\documentclass[twocolumn]{aastex7}
\usepackage[utf8]{inputenc}
\usepackage{amsmath} 
\usepackage{rotating}
\usepackage{newunicodechar}
\newunicodechar{−}{-}
\usepackage{textcomp}
\usepackage{graphicx}
\graphicspath{{figures/}}
\usepackage{lipsum} 
\usepackage{threeparttable}
\usepackage{natbib}
\usepackage{lineno}
\usepackage{comment}
\usepackage{rotating}
\linenumbers
\usepackage{adjustbox}
\usepackage{rotating}
\usepackage{tabularx}

\DeclareRobustCommand{\ion}[2]{%
\relax\ifmmode
\ifx\testbx\f@series
{\mathbf{#1\,\mathsc{#2}}}\else
{\mathrm{#1\,\mathsc{#2}}}\fi
\else\textup{#1\,{\mdseries\textsc{#2}}}%
\fi}

\newcommand{\kms} {\mbox{\,km~s$^{-1}$}}

\newcommand{\HI}{H\,{\sc i}}

\shorttitle{MeerKAT \HI\ in group and field galaxies}
\shortauthors{Cao et al.}

\begin{document}

\title{Massive to Ultra-Diffuse Galaxies in Group and Field Environments with MeerKAT}

\correspondingauthor{Tian-Wen Cao}
\email{twcao@bao.ac.cn}

\author[0000-0002-1335-6212,gname=Tian-Wen,sname=Cao]{Tian-Wen Cao}
\affiliation{National Astronomical Observatories, Chinese Academy of Sciences, Beijing 100101, People$'$s Republic of China}
\affiliation{Guizhou Radio Astronomical Observatory, Guizhou University, Guiyang 550000, People$'$s Republic of China}
\email[]{twcao@bao.ac.cn}

\author[0000-0001-7547-4241,gname=Neeraj,sname=Gupta]{Neeraj Gupta}
\affiliation{Inter-University Centre for Astronomy and Astrophysics, Post Bag 4, Ganeshkhind, Pune 411 007, India; ngupta@iucaa.in}
\email[]{ngupta@iucaa.in}

\author[0000-0003-2658-7893, gname=Francoise,sname=Combes]{Francoise Combes}
\affiliation{Observatoire de Paris, LUX, Coll\`ege de France, PSL University, Sorbonne University, CNRS, Paris, France}
\email[]{francoise.combes@obspm.fr}

\author[0000-0002-3814-9666, gname=Sergei,sname=Balashev]{Sergei A. Balashev}
\affiliation{Independent Researcher}
\email[]{s.balashev@gmail.com}

\author[0009-0002-4905-263X, gname=Pei-bin,sname=Chen]{Pei-bin Chen}
\affiliation{School of Physics and Astronomy, China West Normal University, Nanchong 637009, People$'$s Republic of China}
\affiliation{Local Universe and Time Domain Astronomy Laboratory, China West Normal University, No. 1 Shida Road, Nanchong 637002, People$'$s Republic of China}
\email[]{chenpb@cwnu.edu.cn}

\author[0000-0003-0202-0534, gname=Cheng,sname=Cheng]{Cheng Cheng} 
\affiliation{Chinese Academy of Sciences South America Center for Astronomy, National Astronomical Observatories, CAS, Beijing 100101, People$'$s Republic of China}
\affiliation{Key Laboratory of Optical Astronomy, NAOC, 20A Datun Road, Chaoyang District, Beijing 100101, People$'$s Republic of China}
\email[]{chengcheng@nao.cas.cn}

\author[0000-0002-4641-2532, gname=Venu,sname=Kalari]{Venu Kalari}
\affiliation{Gemini Observatory/NSF's NOIRLab, Casilla 603, La Serena, Chile}
\email[]{venu.kalari@noirlab.edu}

\author[0000-0002-4912-9388, gname=Jens-Kristian,sname=Krogager]{Jens-Kristian Krogager} 
\affiliation{French-Chilean Laboratory for Astronomy, IRL 3386, CNRS and U. de Chile, Casilla 36-D, Santiago, Chile}
\affiliation{Universit{\' e} Lyon 1, ENS de Lyon, CNRS, CRAL, UMR 5574, Saint-Genis-Laval, France}
\email[]{jens-kristian.krogager@univ-lyon1.fr}

\author[0000-0002-5777-1629, gname=Pasquier,sname=Noterdeame]{Pasquier Noterdaeme} 
\affiliation{Institut d'astrophysique de Paris, CNRS-SU, UMR 7095, 98bis bd Arago, 75014 Paris, France}
\email[]{noterdaeme@iap.fr}

\author[0000-0002-2097-7543, gname=Luis Enrique,sname=Perez Montano]{Luis Enrique P\'erez-Monta\~no} 
\affiliation{Department of Astronomy, Xiamen University, 422 Siming South Road, Xiamen 361005, People$'$s Republic of China}
\affiliation{Institute for Astronomy, School of Physics, Zhejiang University, Hangzhou 310027, People$'$s Republic of China}
\email[]{le.perezmontano@zju.edu.cn}

\author[0000-0001-6083-956X, gname=Ming,sname=Zhu]{Ming Zhu} 
\affiliation{National Astronomical Observatories, Chinese Academy of Sciences, Beijing 100101, People$'$s Republic of China}
\affiliation{University of Chinese Academy of Sciences, Beijing 100049, People$'$s Republic of China}
\affiliation{CAS Key Laboratory of FAST, National FAST, National Astronomical Observatories, Chinese Academy of Sciences, Beijing 100101, People$'$s Republic of China}
\affiliation{Guizhou Radio Astronomical Observatory, Guizhou University, Guiyang 550000, People$'$s Republic of China}
\email[]{mz@nao.cas.cn}

\begin{abstract}
We present a study of a galaxy group environment at redshift, $z$ $\sim$ 0.019 $-$ 0.022, using wide-field \HI\ imaging from the MeerKAT Absorption Line Survey (MALS) pointing centered on J012613.24+142013.1.
Six of the eleven galaxies within the group redshift range are resolved in \HI, including one ultra-diffuse galaxy (UDG), together with two foreground galaxies at z$\simeq$0.016. Combining these resolved \HI\ maps with multi-wavelength data, we derive disk inclinations and radial surface-density profiles for eight galaxies spanning massive spirals to ultra-diffuse systems, with tilted-ring kinematic models obtained for seven of these. 
Accounting for the measurement uncertainties, 
maximum azimuthally averaged \HI\ surface densities span approximately 5 -- 20\,M$_\odot$\,pc$^{-2}$.
Among the two UDGs, AGC\,748816 has a centrally peaked \HI\ surface-density profile, while AGC\,110319 shows a tentative central depression.
Most detected galaxies lie above the \HI-selected stellar-mass–\HI-mass relation while having star formation rates near or below the main sequence, indicating gas-rich but inefficient star formation. Most galaxies lie within the scatter of reference baryonic Tully-Fisher relations, and the group UDG AGC\,748816 shows the clearest upward displacement, though residual uncertainties in its inclination and kinematic modelling do not provide compelling evidence for an anomalously low dark-matter content.
\end{abstract}


\keywords{Galaxy groups, Star formation quenching, Dark matter, Ultra-diffuse galaxies}

\section{Introduction}

Galaxy groups are the most common galaxy systems in the local Universe \citep[e.g.,][]{2005MNRAS.362.1233E}.
They occupy an intermediate regime between isolated field galaxies and dense clusters.
Within these systems, gravitational encounters, tidal interactions, mergers, and gas stripping play crucial roles in transforming galaxy morphology, quenching star formation, and redistributing baryons across the group halo. Extensive observational work has shown that galaxy morphology correlates with both group-centric distance and local galaxy density\citep[e.g.,][]{1984ApJ...281...95P, 2001ApJ...549..172T, 2009ApJ...693L..71D, 2012MNRAS.424..232W, 2021A&A...646A.151P, 2025A&A...698A..57L},
and that group environments have a significant impact on the star formation activity of their member galaxies \citep[e.g.,][]{2011MNRAS.412.2303B, 2013MNRAS.434.3089Z, 2016MNRAS.455.4013D, 2018ApJ...857...71B, 2024ApJ...966...18T, 2024A&A...691A.107C}.

Galaxy groups harbor a wide range of galaxy types, ranging from massive spirals and early-type systems to late-type and dwarf systems \citep{2016MNRAS.456.3961S, 2019ApJ...885...88P, 2024ApJ...976L..18P}. Numerous faint and extended systems have been discovered in group environments \citep{2019MNRAS.487.5272J, 2021NatAs...5.1255B, 2023ApJ...946L..37N}, including low surface brightness galaxies \citep[LSBGs;][]{1987AJ.....94...23B, 1992AJ....103.1107S}.

\begin{figure*}[htb]
  \begin{center} 
  \includegraphics[width=6.0in]{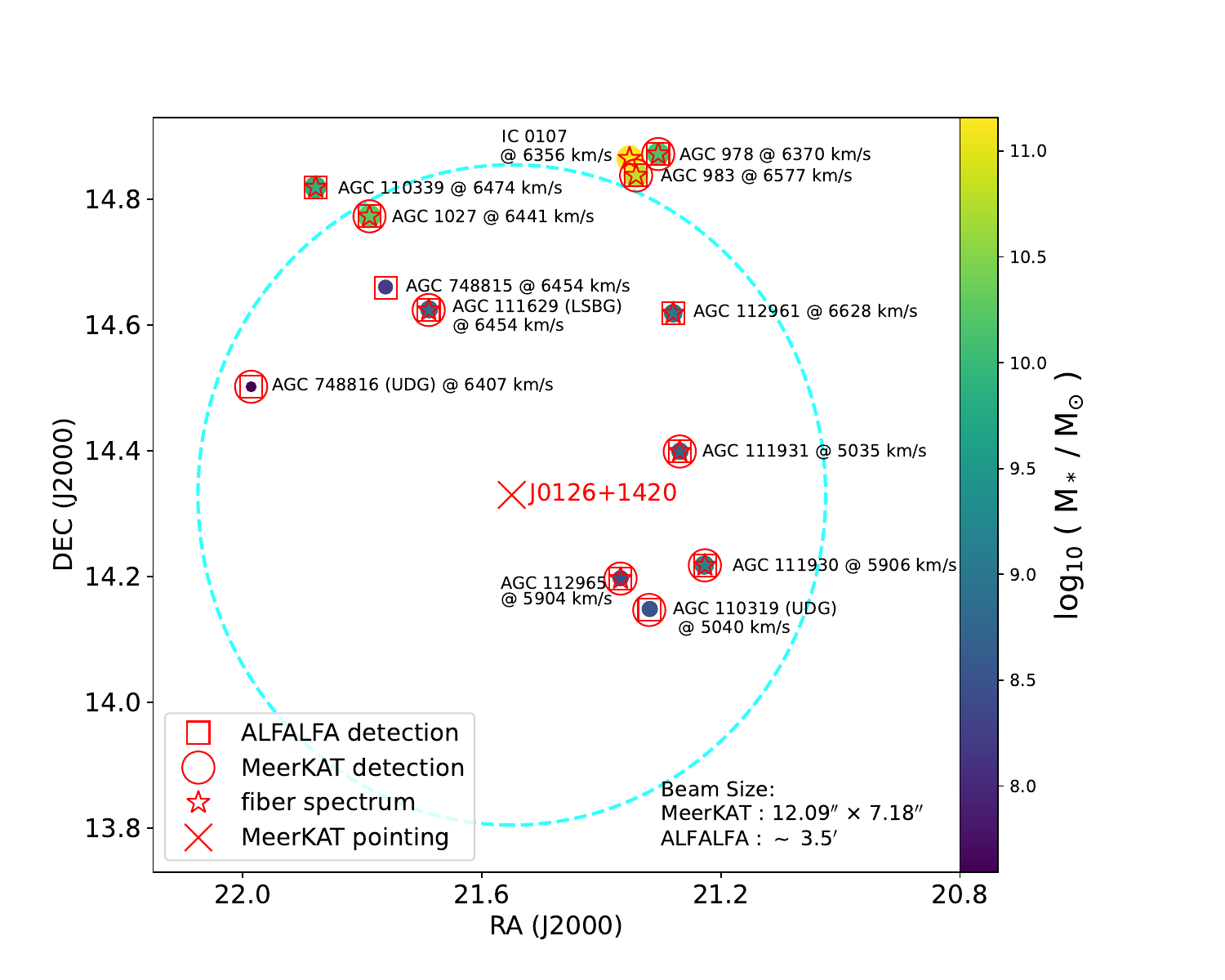}
  \end{center}
\caption{
Spatial distribution of the 13 sample galaxies in the field of MALS J0126$+$1420 (Table~\ref{tab:meerkat}). The dashed cyan circle shows the MeerKAT FWHM (63$\arcmin$), centered on an unrelated radio source (red $\times$ marker). 
Red squares and circles indicate galaxies with ALFALFA and spatially resolved \HI\ detections from MeerKAT, respectively. Of the nine MeerKAT detections, eight have spatially resolved \HI\ maps, and the remaining one (AGC\,112965) was detected at low S/N through a targeted search (see text for details).
Red stars denote galaxies with optical fiber spectra. The color of each symbol represents stellar mass estimated using MLCRs (Section~\ref{sec:mstar}).
\label{DSP}}
\end{figure*}

LSBGs, defined by their diffuse stellar disks and central surface brightness ($\mu_0$) fainter than 23.0 mag arcsec$^{-2}$ in the B-band \citep{1997ARA&A..35..267I}, or equivalently $\mu_{0,g}>22.5$\,mag\,arcsec$^{-2}$ in the $g$-band \citep{2016MNRAS.456.1607D} for typical LSBG colours, are significantly dominated by dark matter \citep[DM;][]{1995MNRAS.273L..35Z, 2001AJ....122.2396D, 2002A&A...385..816D} with a remarkably inefficient star formation \citep{2009ApJ...696.1834W, 2017AJ....154..116C, 2024Univ...10..432C}. Ultra-diffuse galaxies (UDGs) identified more recently, are extremely low surface-brightness systems with $\mu_0$ fainter than 24.0 mag arcsec$^{-2}$ and effective radii (R$_e$) larger than 1.5\,kpc in $g$-band \citep{2015ApJ...798L..45V, 2016ApJ...833..168M}. Although UDGs were first extensively studied in rich clusters, subsequent work has shown that they are also found in lower-density environments, including galaxy groups, the field, and even void-like regions \citep{2017MNRAS.468.4039R, 2019MNRAS.486..823R, 2020ApJS..247...46B}. These systems therefore provide a useful laboratory for testing whether diffuse galaxies are produced mainly by internal processes, such as high halo spin and stellar feedback, or by environmental mechanisms, such as tidal interactions, gas stripping, and group preprocessing.

\cite{2020AJ....159..138D} show that most \HI-selected LSBGs share similar stellar mass-to-light ratio (M$_*$/L) to those of normal galaxies and \cite{2022ApJ...941...11K} find that the (M$_*$/L) in $g$-band of \HI-selected UDGs is not offset from their dwarf sample. However, some dynamical mass-to-light ratio (M$_d$/L) outlier UDGs are reported in the literature, for example,  VCC 1287 in the Virgo cluster \citep[][]{2020MNRAS.495.2582G}, Dragonfly 44 \citep{2016ApJ...828L...6V, 2019ApJ...880...91V} and NGC\,1052-DF2 \citep{2018Natur.555..629V}. 
In addition, resolved \HI\ kinematics of gas-rich UDGs have revealed systems with unexpectedly low rotation velocities, offsets from the baryonic Tully-Fisher relation (BTFR), low-density or low-concentration halos, and in some cases apparently baryon-dominated inner regions \citep{2019MNRAS.488.3222S, 2019ApJ...883L..33M, 2021ApJ...909...20S, 2022MNRAS.512.3230M, 2024A&A...689A.344M, 2024A&A...692A.217S}.
These galaxies challenge the conventional picture of galaxy formation and evolution, particularly in understanding how such diffuse systems with low-mass dark matter halos can survive and evolve within dense environments. The coexistence of massive, normal, and diffuse galaxies within the same group environment provides an opportunity to explore how environmental processes shape galaxies across a wide range of stellar masses, surface brightnesses, and gas content.

The formation, environmental dependence, and internal properties of low surface brightness systems have been extensively studied through numerical simulations. Simulations have investigated the role of large halo spin, feedback-regulated star formation, and angular momentum redistribution in shaping their diffuse stellar distributions \citep{2021MNRAS.502.4262J, 2022MNRAS.514.5840P, 2025A&A...701A.272S}. These studies generally suggest that low surface brightness systems preferentially form in dark matter halos with elevated spin parameters \citep{2024MNRAS.533...93P} and highlight the importance of environmental processes, such as tidal heating, ram-pressure stripping, and galaxy–galaxy interactions in group and cluster environments  \citep{2018MNRAS.480L.106O, 2019MNRAS.485..796M}.

Neutral hydrogen (\HI) plays a central role in this context, serving as a sensitive tracer of environmental influence. As the most extended baryonic component in galaxies, the \HI\ gas is often the first to be influenced by environmental processes such as tidal stripping, ram-pressure, or gas accretion from the intra-group medium. 
Large-scale \HI\ surveys, such as Arecibo Legacy Fast ALFA \citep[ALFALFA;][]{2005AJ....130.2598G}, the \HI\ Parkes All-Sky Survey \citep[HIPASS;][]{2001MNRAS.322..486B}, and the FAST all sky \HI\ survey \citep[FASHI;][]{2024SCPMA..6719511Z, 2026SCPMA..6929811Z}, have provided broad statistical insights into the gas content of galaxies in various environments. However, their coarse spatial resolution limits detailed morphological and kinematic analysis.
With the advent of MeerKAT \citep{2016mks..confE...1J}, an SKA-mid precursor, and ASKAP \citep[][]{Hotan2021} telescopes, high-sensitivity and high-angular-resolution \HI\ imaging of large samples of galaxy groups is now feasible. 
Through large field-of-views and unprecedented surface-brightness sensitivity these are enabling  detections of diffuse tidal gas, bridges, and low-column-density features that trace interactions and assembly in a wide range of galaxy environments \citep[e.g.,][]{Serra2015, Westmeier2022, 2024A&A...688A.109D, 2025A&A...701A..86L, 2025ApJ...984L..46S, 2026A&A...708A.122A}.

The MeerKAT Absorption Line Survey \citep[MALS;][]{Gupta17mals} is a recently completed large programme at the MeerKAT telescope. The survey covers about 400 pointings centered at radio sources brighter than 200\,mJy at 1\,GHz in the $L$-band (900 - 1670\,MHz) and the UHF band (580 - 1015\,MHz).
\citet[][]{Gupta21salt} described the target selection process optimized for the requirements of the extragalactic \HI\ 21-cm and OH 18-cm absorption line search, and ensuring reasonable observability with the MeerKAT telescope in the southern hemisphere \citep[see also][]{Krogager19}. The sky coverage of 391 pointings observed at L-band between April 1$^{\rm st}$, 2020 and January 18$^{\rm th}$, 2021 is described in \citet[][]{Deka24dr1}.
MeerKAT telescope's large field of view with a full width at half maximum (FWHM) of $\theta\approx88\arcmin$ at 1\,GHz makes it a highly competitive survey for \HI\ emission line \citep[e.g.,][]{Maina2022, Boettcher21, depalma2025}, radio-continuum from star-forming and active galaxies \citep[e.g.,][]{Deka24dr1, Wagenveld2024} and Galactic \HI\ absorption \citep[][]{Gupta2025galhi}.

In this paper, we present a detailed \HI\ study of galaxies in the field of MALS pointing J012613.24+142013.1, hereafter referred to as J0126+1420. This pointing was selected from the full MALS survey based on the availability of extensive ancillary data and previous studies of the associated galaxy group at redshift $z\sim$0.02 \citep{2025AJ....170..296C}. This provides a well-characterized environmental context for our analysis. We focus on the frequency range covering 1379.6--1400.5\,MHz, corresponding to the \HI\ 21-cm line over the redshift range $z\sim 0.014–0.030$, which encompasses all galaxies studied here (Figure~\ref{DSP}). We note that the MALS pointing is centered on a background radio source (577\,mJy, $z$ = 0.573) targeted for H\,{\sc i} and OH absorption line searches, unrelated to the galaxies discussed here.
The paper is organized as follows.
In Section~\ref{sec:sample} we describe the MeerKAT and other multi-wavelength datasets used in this study. Observational results are presented in Section~3. We discuss our results in Section~4 and summarize our conclusions in Section~5. All magnitudes in this paper are AB magnitudes. We adopt a $\Lambda$CDM cosmology with $\rm \Omega_{m}$ = 0.3, $\rm \Omega_{\Lambda}$ = 0.7, $\rm H_0$ = 70\,km\,s$^{-1}$\,Mpc$^{-1}$.

\section{Sample and observational data}
\label{sec:sample}

\begin{sidewaystable*}
\caption{Properties of the sample galaxies in the field of MALS J0126+1420.}
\centering
\footnotesize
\setlength{\tabcolsep}{2.5pt}
\begin{tabular}{c c c c c c c c c c c c c c c c}
\hline\hline
1 & 2 & 3 & 4 & 5 & 6 & 7 & 8 & 9 & 10 & 11 & 12 & 13 & 14 & 15 & 16 \\
\hline
Source & z & D & M$_g$ & (g$-$r) & log & SFR & [M/H] & W$_{50}$ & \HI\ peak & \HI\ flux & log\,\HI\ mass & SFR & $i_{\textsc{HI}}$ & PA & V$_{circ}$ \\
name & & & & & & NUV$_{corr}$ & & & & & & 1.4\,GHz & & & \\
& & [Mpc] & [mag] & [mag] & [M$_*$/M$_{\odot}$] & [M$_{\odot}$\,yr$^{-1}$] & & [km\,s$^{-1}$] & [km\,s$^{-1}$] & [Jy\,km\,s$^{-1}$] & [M$_{\textsc{Hi}}$/M$_{\odot}$] & [M$_{\odot}$\,yr$^{-1}$] & [deg] & [deg] & [km\,s$^{-1}$] \\
\hline
AGC\,112961$^{\dag}$ & 0.022 & 90.3 & $-$17.15 & 0.540 & 9.02 & -- & -- & 154$^{*}$ & -- & $<$0.14 & $<$8.43 & -- & -- & -- & -- \\
 & & $\pm$2.3 & $\pm$0.06 & $\pm$0.005 & $\pm$0.02 & & & & & & & & & & \\
AGC\,983 & 0.022 & 89.6 & $-$20.23 & 0.800 & 10.80 & 0.36 & 0.02 & 431.4 & 6374.6 & 7.87 & 10.17 & 1.24 & 73 & 109 & 226 \\
(AGN) & & $\pm$2.2 & $\pm$0.05 & $\pm$0.001 & $\pm$0.02 & $\pm$0.02 & & $\pm$10.0 & $\pm$3.6 & $\pm$0.47 & $\pm$0.03 & $\pm$0.38 & $\pm$5 & & $^{+16}_{-13}$ \\
IC\,0107 & 0.0212 & 86.4 & $-$21.08 & 0.820 & 11.12 & -- & -- & -- & -- & $<$0.3 & $<$8.7 & -- & -- & -- & -- \\
(AGN) & & $\pm$6.0 & $\pm$0.15 & $\pm$0.001 & $\pm$0.06 & & & & & & & & & & \\
AGC\,978 & 0.0212 & 86.6 & $-$19.93 & 0.540 & 10.13 & 0.84 & 0.04 & 145.6 & 6299.9 & 2.97 & 9.72 & 1.50 & 42 & 98 & 103 \\
 & & $\pm$2.2 & $\pm$0.06 & $\pm$0.001 & $\pm$0.02 & $\pm$0.05 & & $\pm$6.2 & $\pm$2.4 & $\pm$0.30 & $\pm$0.05 & $\pm$0.46 & $\pm$5 & & $^{+15}_{-12}$ \\
AGC\,1027 & 0.0214 & 87.6 & $-$20.65 & 0.460 & 10.26 & 2.66 & $-$0.02 & 256.0 & 6363.1 & 5.58 & 10.00 & 4.21 & 40 & 352 & 206 \\
 & & $\pm$2.2 & $\pm$0.05 & $\pm$0.001 & $\pm$0.02 & $\pm$0.17 & & $\pm$10.5 & $\pm$2.4 & $\pm$0.27 & $\pm$0.03 & $\pm$1.28 & $\pm$5 & & $^{+33}_{-39}$ \\
AGC\,110339$^{\dag}$ & 0.0216 & 88.1 & $-$19.25 & 0.540 & 9.86 & 0.77 & $-$0.05 & 81$^{*}$ & -- & $<$0.23 & $<$8.62 & -- & -- & -- & -- \\
 & & $\pm$2.4 & $\pm$0.06 & $\pm$0.001 & $\pm$0.02 & $\pm$0.05 & & & & & & & & & \\
AGC\,111629 & 0.0215 & 87.8 & $-$17.47 & 0.390 & 8.84 & 0.11 & $-$1.63 & 188.8 & 6541.5 & 1.78 & 9.50 & -- & 72 & 21 & 87$^*$ \\
(LSBG; AGN) & & $\pm$2.1 & $\pm$0.05 & $\pm$0.005 & $\pm$0.02 & $\pm$0.01 & & $\pm$6.0 & $\pm$2.8 & $\pm$0.15 & $\pm$0.05 & & $\pm$5 & & $^{+7}_{-6}$ \\
AGC\,748815$^{\dag}$ & 0.0216 & 88.0 & $-$16.20 & 0.330 & 8.21 & -- & -- & 144$^{*}$ & -- & $<$0.05 & $<$7.96 & -- & -- & -- & -- \\
 & & $\pm$2.3 & $\pm$0.06 & $\pm$0.010 & $\pm$0.03 & & & & & & & & & & \\
AGC\,748816 & 0.0213 & 87.2 & $-$16.53 & $-$0.04 & 7.59 & -- & -- & 76.2 & 6426.4 & 1.63 & 9.47 & -- & 29 & 56 & 68$^*$ \\
(UDG) & & $\pm$2.2 & $\pm$0.06 & $\pm$0.01 & $\pm$0.03 & & & $\pm$5.2 & $\pm$2.0 & $\pm$0.22 & $\pm$0.06 & & $\pm$5 & & $^{+20}_{-25}$ \\
AGC\,111930 & 0.0197 & 82.5 & $-$18.33 & 0.400 & 9.21 & 0.21 & $-$1.06 & 177.3 & 5892.4 & 1.78 & 9.45 & 0.12 & 57 & 297 & 96$^{*}$ \\
(AGN) & & $\pm$2.3 & $\pm$0.06 & $\pm$0.002 & $\pm$0.03 & $\pm$0.01 & & $\pm$7.4 & $\pm$2.6 & $\pm$0.18 & $\pm$0.05 & $\pm$0.04 & $\pm$5 & & $^{+10}_{-13}$ \\
AGC\,112965$^{\ddag}$ & 0.0197 & 82.5 & $-$16.66 & 0.360 & 8.46 & 0.07 & $-$0.31 & 88 & -- & 0.72 & 9.06 & -- & -- & -- & -- \\
 & & $\pm$2.2 & $\pm$0.06 & $\pm$0.008 & $\pm$0.03 & $\pm$0.01 & & $\pm$6.6 & & $\pm$0.11 & $\pm$0.07 & & & & \\
AGC\,111931 & 0.0168 & 69.8 & $-$17.57 & 0.310 & 8.72 & 0.12 & 0.01 & 183.7 & 4960.9 & 3.81 & 9.64 & 0.009 & 71 & 100 & 102 \\
 & & $\pm$2.4 & $\pm$0.07 & $\pm$0.003 & $\pm$0.03 & $\pm$0.01 & & $\pm$5.7 & $\pm$1.9 & $\pm$0.21 & $\pm$0.04 & $\pm$0.006 & $\pm$5 & & $^{+14}_{-12}$ \\
AGC\,110319 & 0.0168 & 69.9 & $-$15.69 & 0.570 & 8.50 & 0.09 & -- & 20.8 & 5040.7 & 1.74 & 9.30 & -- & 31 & 297 & -- \\
(UDG) & & $\pm$2.3 & $\pm$0.07 & $\pm$0.010 & $\pm$0.04 & $\pm$0.01 & & $\pm$6.0 & $\pm$0.9 & $\pm$0.15 & $\pm$0.05 & & $\pm$14 & & \\
\hline
\end{tabular}
\normalsize
\begin{tablenotes}
\small
\item Column 1: Source names from ALFALFA or the Index Catalog \citep{1895MmRAS..51..185D}. Targets marked $^{\dag}$ are ALFALFA-detected but MeerKAT-undetected; targets marked $^{\ddag}$ were detected by MeerKAT at insufficient S/N for morphological analysis.
Column 2: Redshift based on the heliocentric recession velocity. The values are taken from the ALFALFA catalog \citep{2018ApJ...861...49H}, except for IC\,0107, whose redshift is from \citet{2003AJ....126.2268W}.
Column 3: Distances from ALFALFA \citep{2018ApJ...861...49H}; for IC\,0107, we follow the same methodology.
Column 4: Absolute $g$-band magnitude from DECaLS.
Column 5: ($g - r$) color from DECaLS $g-$band and $r-$band photometry.
Column 6: Stellar mass derived using the MLCR method (see Section~\ref{sec:mstar}).
Column 7: SFR from \citet{2020AJ....160..271D} estimated from the NUV band luminosity with dust extinction correction based on mid-IR flux at 22\,$\mu$m, adopting a \citet{1955ApJ...121..161S} IMF (see Section~\ref{sec:mstar}).
Column 8: Light-weighted mean metallicity ([M/H]) obtained using the pPXF method (see Section~\ref{sec:mstar}). The typical uncertainty on [M/H] is 0.1--0.2\,dex \citep{2005MNRAS.362...41G}.
Column 9: the global \HI\ profile width measured at 50\% of the corresponding
peak flux on each side of the profile (W$_{50}$). For AGC\,112961, AGC\,110339, and AGC\,748815, the W$_{50}$ values used to estimate the upper limits are based on FASHI measurements (private communication); these are marked with $^{*}$. All three are weakly detected by FASHI but undetected by MeerKAT.
Column 10: Heliocentric velocity of the \HI\ peak and its $1\sigma$ uncertainty. Blank for galaxies not detected by MeerKAT and for AGC\,112965.
Column 11: Integrated \HI\ flux density from MeerKAT observations. For sources not detected by SoFiA, $3\sigma$ upper limits are given assuming W$_{50}$.
Column 12: \HI\ gas mass derived from the MeerKAT \HI\ flux. For non-detections, upper limits are given.
Column 13: Star formation rate estimated from the 1.4\,GHz radio continuum from MALS DR1 \citep{Deka24dr1}.
Column 14: Morphological inclination inferred from the beam-corrected \HI\ moment-0 distribution at native resolution (see Section~\ref{sec:geometry}). For AGC\,110319, the inclination is obtained from the separate global disk analysis.
Column 15: Position angle (PA) determined at native resolution and adopted as a modelling input (see Section~\ref{sec:modle}) and held fixed in the kinematic fits; therefore, no independent formal uncertainty is quoted.
Column 16: Characteristic circular velocity measured from the asymmetric-drift-corrected \textsc{3DBarolo} rotation curve fitted to the cube smoothed to a circular $15''$ FWHM beam. For AGC\,111629, AGC\,748816, and AGC\,111930, which were fitted with only two rings, the values marked with $^{*}$ are fitted outer-ring rotation velocities. No velocity is assigned to AGC\,110319.
\end{tablenotes}
\label{tab:meerkat}
\end{sidewaystable*}

Our sample consists of 13 galaxies in the field of J0126+1420 -- eleven at $z \sim 0.02$ that form a group and two foreground galaxies at $z \sim 0.016$. Their properties are listed in Table~\ref{tab:meerkat}. Twelve have \HI\ detections from ALFALFA \citep{2018ApJ...861...49H} and/or the MeerKAT observations presented here (Figure~\ref{DSP}). The remaining galaxy IC\,0107 at $z\sim0.02$ is not detected in \HI\ and is included on the basis of its optical redshift derived from the Redshift-distance Survey of Nearby Early-type Galaxies (ENEAR, \citealt{2003AJ....126.2268W}). Of the nine galaxies detected by MeerKAT, eight have spatially resolved \HI\ maps, as described below. One additional galaxy, AGC\,112965, was detected through a targeted search but at insufficient S/N for morphological analysis.

\subsection{Distribution of Galaxies}

Figure~\ref{DSP} shows the spatial distribution of all 13 galaxies, with the eleven group members at $z \sim 0.02$ and the two foreground galaxies, AGC\,111931 and AGC\,110319, at $z \sim 0.016$. 
Among the galaxies shown in Figure~\ref{DSP}, three systems -- AGC\,111629, AGC\,748816, and AGC\,110319 -- are classified as LSBGs, and the latter two additionally satisfy the adopted UDG criteria. The $g$-band central surface brightnesses and effective radii are adopted from the published photometric analyses cited below. Following the classification scheme adopted in this work, galaxies with a central surface brightness of $\mu_{0,g}>22.5$\,mag\,arcsec$^{-2}$ are classified as LSBGs, while UDGs are required to have both $\mu_{0,g}>24.0$ mag arcsec$^{-2}$ and an effective radius of $R_e>1.5$ kpc.
AGC\,111629 is a blue, edge-on LSBG with an inclination-corrected $\mu_{0,g}^{face-on}=22.7$\,mag\,arcsec$^{-2}$ \citep{2023ApJ...948...96C}. Integral-field spectroscopic observations of its ionized gas were subsequently presented by \citet{2025AJ....170..296C}. AGC\,748816 is an \HI-bearing UDG identified by \citet{2017ApJ...842..133L}, with published values of $\mu_{0,g}=24.4$\,mag\,arcsec$^{-2}$ and $R_e=3.4$ kpc \citep{2019MNRAS.490..566J}. The foreground galaxy AGC\,110319 was classified as an LSBG by \citet{2015AJ....149..199D} and satisfies the adopted UDG criteria, with $\mu_{0,g}=24.3$ mag arcsec$^{-2}$ and $R_e=2.7$ kpc \citep{2018ApJS..235...18L}.

\subsection{The MeerKAT observations and data reduction}
\label{sec:mrkt}
%

The pointing centered at RA: {\tt 01:26:13.24} and Dec: {\tt +14.20.13.1} was observed for 56 minutes on 2020 June 14, using 59 antennas of the MeerKAT-64 array. The total observing time was split into three scans of 1120\,s,  each at a different hour angle to improve the uv-coverage. The total 856\,MHz bandwidth was centered at 1283.9869\,MHz and split into 32768 frequency channels, with a channel spacing of 26.123\,kHz, corresponding to  5.5\,\kms\ at the rest frequency of the \HI\ 21-cm line. 
The correlator dump time was 8\,s. For dual, linearly polarized L-band feeds with orthogonal polarizations labeled X and Y, the data were acquired for all four polarization products: XX, XY, YX, and YY. 

The MALS data were processed using the Automated Radio Telescope Imaging Pipeline (ARTIP) based on NRAO's Common Astronomy Software Applications (CASA) package \citep[][]{Casa22}.  The details of calibration and spectral line imaging using ARTIP are provided in \citet[][]{Gupta21}. 
For spectral line processing, {\tt ARTIP} partitions the frequency band into 15 spectral windows (SPW, 0 to 14). For the galaxy group presented here, we used the frequency range 1379.6--1400.5\,MHz from SPW-9 covering 1351--1411\,MHz.  The calibrated and continuum-subtracted visibilities, with {\tt ROBUST = 0} weighting, were inverted to obtain the dirty image cube.  
The restored synthesized beam size for the \HI\ cube is $12\farcs1$ $\times$ $7\farcs2$ with a position angle of $-4\fdg3$, and the rms noise across the cube is 0.8\,mJy\,beam$^{-1}$ per 5.5\,\kms\ channel.

We used the 3D source-finding application SoFiA\,2 \citep{serra2015sofia, Westmeier2021}\footnote{\href{https://gitlab.com/SoFiA-Admin/SoFiA-2}{https://gitlab.com/SoFiA-Admin/SoFiA-2}} to identify the voxels with \HI\ emission signal in the cube. The SoFiA mask was used to iteratively deconvolve the line emission down to the single-channel rms \citep[see][for details]{Boettcher21}. We explored a range of SoFiA source-finding parameters, including detection thresholds and spatial and spectral kernel sizes, to assess the reliability of detections and ensure completeness. To place the non-detections on a firm footing, we ensured that the clean mask encompassed the optical positions of all the galaxies, including those not detected in \HI, so that  spectral extraction is performed on a consistently deconvolved cube. The deconvolved image cube was finally passed through SoFiA to catalog the galaxies and generate \HI\ moment maps and the integrated spectra. The correction for the primary beam attenuation in the range 0.4 -- 0.8 was applied using the \texttt{katbeam} model \citep{Mauch20}.
%

\begin{figure}[!ht]
  \begin{center}
  \vbox{
  \includegraphics[width=\columnwidth]{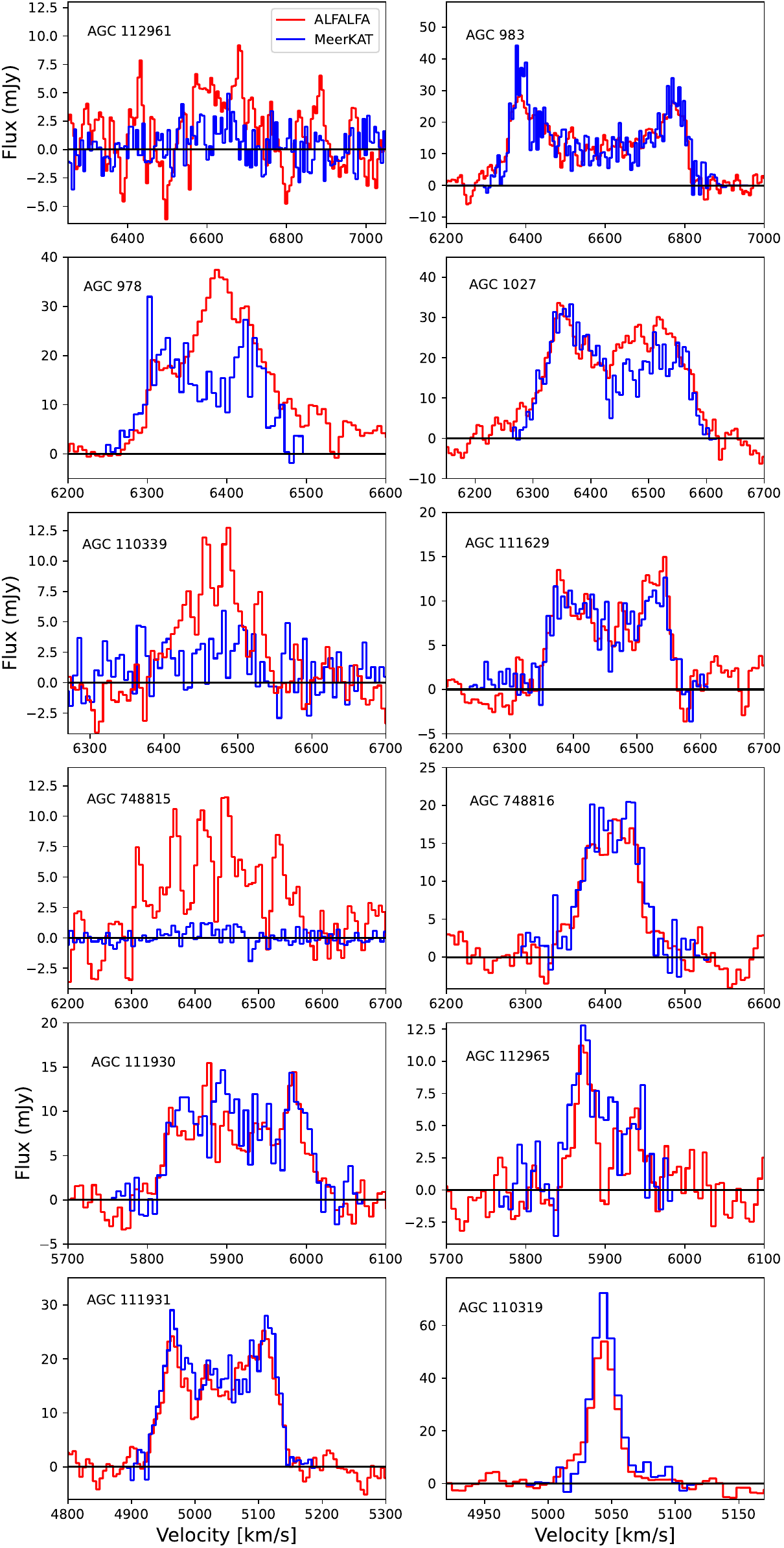}
  }
  \end{center} 
  \caption{The global \HI\ profiles of the 12 ALFALFA-detected galaxies (red line), with recession velocities on the $x$-axis given in the Heliocentric frame. Of these, eight also have spatially resolved \HI\ detections from MeerKAT (blue line). AGC\,111931 and AGC\,110319 in the last two panels are foreground galaxies. IC\,0107 is not shown because it is undetected in both MeerKAT and ALFALFA.}
\label{spectra}
\end{figure}

The \HI\ spectra from MeerKAT and ALFALFA are shown in Figure~\ref{spectra} for all 12 ALFALFA-detected  galaxies (see also Tables~\ref{tab:meerkat} and \ref{tab:additional_data}). IC\,0107 is not included as it is undetected in both surveys.
Of the nine galaxies detected by MeerKAT, eight have spatially resolved \HI\ maps, detected through a blind search. Their morphologies and kinematics are discussed in Section~\ref{sec:group}. For the moment-1 and moment-2 maps, we used only pixels where the signal exceeds 1.5 times the local rms. The ninth, AGC\,112965, was recovered through a targeted search at its optical position but at insufficient S/N for morphological analysis. 

For the four galaxies namely AGC\,112961, IC\,0107, AGC\,110339, and AGC\,748815,  undetected by SoFiA, we extracted spectra at the optical positions. Of these, AGC\,112961, AGC\,110339 and AGC\,748815 have ALFALFA detections that are not recovered in the MeerKAT observations. 
For AGC\,748815 in particular, the nearest MeerKAT detection (AGC\,111629) lies at only 4.7$\arcmin$ with $\Delta V$ = 66\,\kms\ (see Section~\ref{sec:group}), suggesting that the ALFALFA detection may reflect confusion with extended \HI\ emission from AGC\,111629. 
Weak \HI\ emission is detected in the FASHI spectra of AGC\,112961, AGC\,110339, and AGC\,748815 (private communication), at fluxes below the ALFALFA values, suggesting that the ALFALFA fluxes are overestimated, possibly due to confusion, baseline uncertainties, or RFI. The MeerKAT non-detections further suggest that the \HI\ in these galaxies has low column density, falling below our per-channel sensitivity. 
Note that the MeerKAT upper limits in Table~\ref{tab:meerkat} apply to emission within the extraction aperture at the optical position. All four galaxies are treated as MeerKAT non-detections for the remainder of this analysis.

%

\subsection{Multi-wavelength archival data}
\label{sec:multi}

Ten galaxies at redshift $\sim$ 0.02, excluding IC\,0107, and two foreground galaxies within the MALS pointing have single-dish \HI\ 21-cm line measurements from the ALFALFA Extragalactic \HI\ Source Catalog \citep[][see Figure~\ref{DSP} for details]{2018ApJ...861...49H}. The ALFALFA \HI\ fluxes are provided in Table~\ref{app:additional_data}.
Further, of the 13 galaxies, ten, marked by stars in Figure~\ref{DSP}, have optical 3$''$ fiber spectra from SDSS. IC\,0107 has a spectrum from \cite{2003AJ....126.2268W}. 
The distance of IC\,0107 is estimated following the prescription adopted for the ALFALFA catalog and described by \citet{2018ApJ...861...49H}. Specifically, its systemic velocity was converted to the cosmic microwave background reference frame. Since $cz_{\rm CMB}=6047$ km s$^{-1}$ exceeds 6000 km s$^{-1}$, we adopted a pure Hubble-flow distance, $D=cz_{\rm CMB}/H_0$, assuming $H_0=70$ km s$^{-1}$ Mpc$^{-1}$. This yields a distance of 86.4\,Mpc. For consistency, the same ALFALFA distance prescription has been adopted for all the galaxies listed in Table~\ref{tab:meerkat}.

The ultraviolet (UV) imaging data in far- and near-UV, i.e., FUV and NUV, at a spatial resolution of 4$\farcs$2 and 5$\farcs$3, respectively, are available from the Galaxy Evolution Explorer (GALEX, \citealt{2005ApJ...619L...1M}).
The optical imaging data are from the Dark Energy Camera Legacy Survey \citep[DECaLS;][]{2019AJ....157..168D}, including $g$, $r$, and $z$ bands. The median point spread function (PSF) of the $r$-band is approximately 1$\farcs$2 in DECaLS. 
The $g$- and $r$-band photometry of all sample galaxies was measured from the DECaLS images using \textsc{SExtractor} \citep{1996A&AS..117..393B}, from which the absolute magnitude ($M_g$) and ($g-r$) colour listed in Table~\ref{tab:meerkat} were derived.
The star formation rates (SFR) listed in Table~\ref{tab:meerkat} are from \cite{2020AJ....160..271D}, who estimated these using the NUV band, with dust extinction correction based on mid-IR flux at 22\,$\mu$m, adopting a \citet{1955ApJ...121..161S} IMF.

\section{Observational results and analysis}

\subsection{Local density}
\label{sec:density}

To quantify the local three-dimensional environment of each galaxy in our sample, we compute the volume density to the 5th nearest neighbor ($\rho_5$), using galaxies within a line-of-sight velocity window of $\Delta v = \pm 1000$ km s$^{-1}$. The velocity cut serves as a pre-filter to exclude interlopers \citep{2003ApJ...585L...5H, 2015MNRAS.451..660E, 2017ApJ...837...16D}. This estimator is less susceptible to projection effects than measures based only on surface density. 
For estimating $\rho_5$, we construct a combined catalog by merging the ALFALFA $\alpha.100$ catalog \citep[][]{2018ApJ...861...49H} and the NASA$-$Sloan Atlas \citep[NSA v1.0.1;][]{2011AJ....142...31B}, where \HI\ detections provide distances and gas masses for gas-rich systems, while NSA spectroscopic redshifts extend coverage to gas-poor galaxies within the same volume. ALFALFA distances, taken from \citet{2018ApJ...861...49H}, are based on the same prescription described in Section~\ref{sec:multi}. For galaxies lacking ALFALFA detections, distances are computed from NSA spectroscopic redshifts using the same Hubble-flow prescription.

To calculate $\rho_5$ for a given galaxy, we first identify all neighboring galaxies within the velocity window from our combined catalog, then compute the three-dimensional Euclidean distances using
\begin{equation}
r_{ij} = \sqrt{d_i^2 + d_j^2 - 2d_i d_j \cos\theta_{ij}},
\end{equation}
where $d_i$ and $d_j$ are the distances of the two galaxies in Mpc, and $\theta_{ij}$ is their angular separation on the sky in radians \citep{VectorCalculus}. Sorting by $r_{ij}$, we determine $r_5$, i.e., the distance to the 5th nearest neighbor, and compute the volume density as
\begin{equation}
\rho_5 = \frac{5}{\frac{4}{3}\pi (r_5)^3} \quad \text{[Mpc}^{-3}\text{]}.
\end{equation}
The derived $\rho_5$ values are presented in Table~\ref{tab:additional_data}.
The galaxies at z$\sim$0.02 have systematically higher $\rho_5$ values (0.10--0.44\,Mpc$^{-3}$) than the two foreground galaxies ($\sim$0.03\,Mpc$^{-3}$), indicating a relative environmental contrast within our sample.
Given the typical distance uncertainty of $\sim$2~Mpc (Table~\ref{tab:meerkat}, column\,2), $\rho_5$ should be regarded as a qualitative indicator of relative environment rather than a measure of absolute density.
%

\begin{turnpage}
\begin{figure*}
   \centering
   \includegraphics[width=1.1\textwidth]{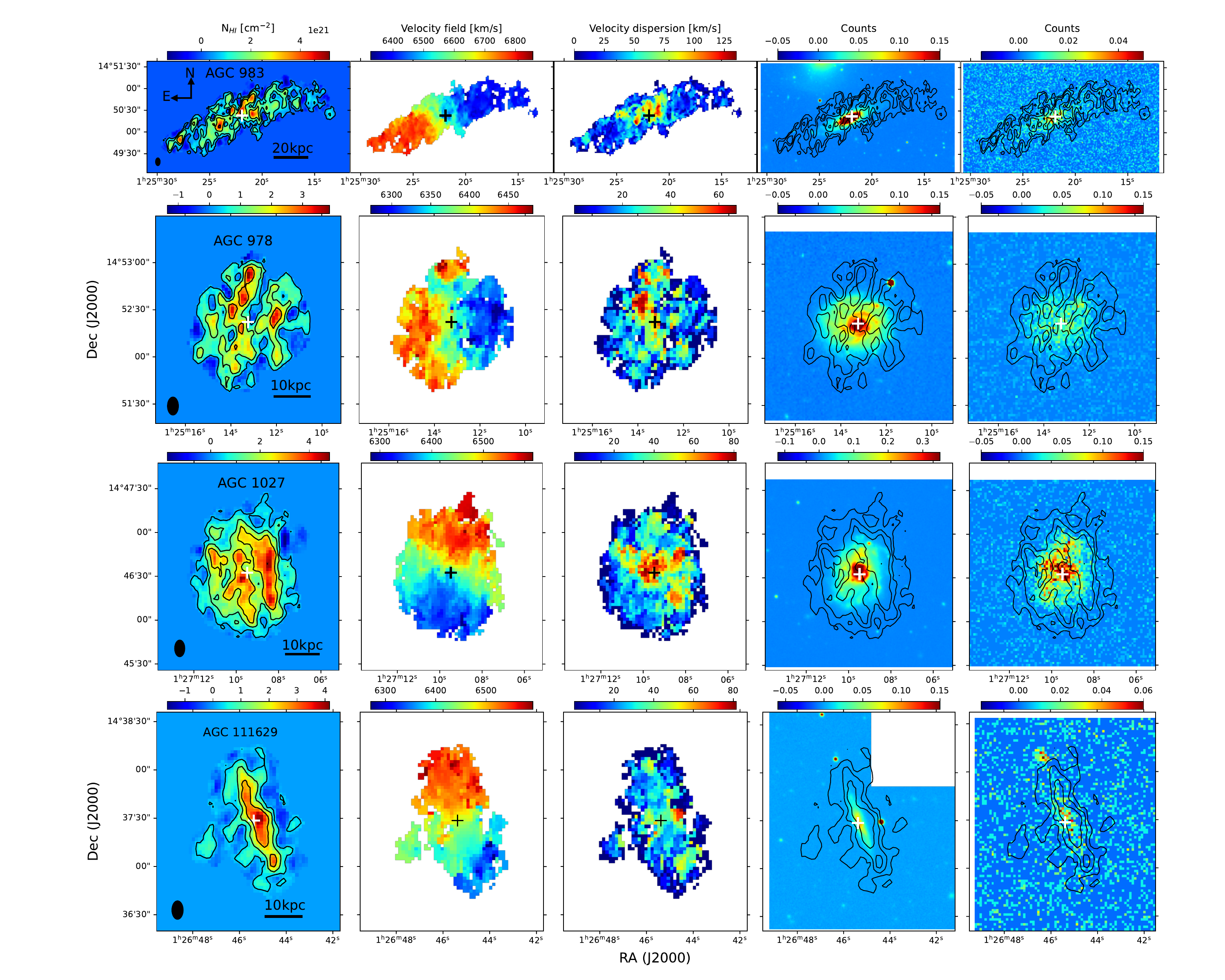}
   \caption{For each galaxy, the five columns show the moment~0 (integrated \HI) map, moment~1 (velocity field), moment~2 (velocity dispersion), DECaLS \textit{g}-band, and GALEX NUV-band images, respectively. The \HI\ column density contour levels are $2^n \times 3.0 \times 10^{20}$\,cm$^{-2}$ ($n=0,\,2,\,3,\,4,\,5$). The outermost contour corresponds approximately to the \HI\ column-density sensitivity, 3.0 $\times 10^{20}$\,cm$^{-2}$ (see Section~\ref{sec:meerkathi}).
   The black-filled circle in the lower-left corner of each moment~0 map represents the MeerKAT synthesized beam.
   The cross marks the \HI\ flux-weighted centre adopted for each galaxy (see Table~\ref{tab:additional_data}).}
   \label{image1}
\end{figure*}
\end{turnpage}
%
\begin{turnpage}
\begin{figure*}[!t]
  \figurenum{3}
  \renewcommand{\theHfigure}{3continued}
  \centering
  \includegraphics[width=1.1\textwidth]{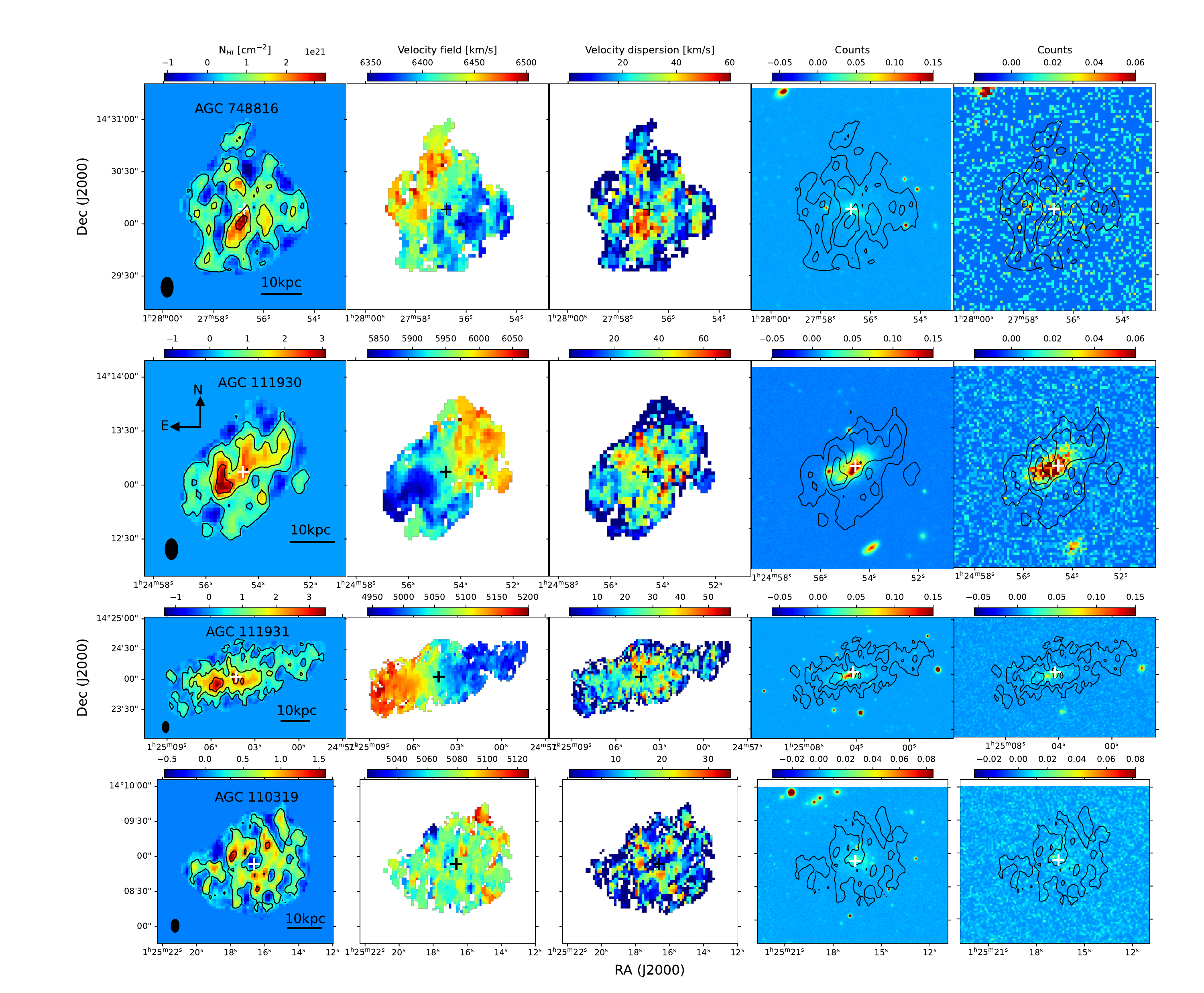}
  \caption{Continued. AGC\,111931 and AGC\,110319 in the last two panels are foreground galaxies.}
  \label{image1b}
\end{figure*}
\end{turnpage}

\subsection{Stellar mass and stellar metallicity}
\label{sec:mstar}

We apply the mass-to-light versus colour relations (MLCRs) of \cite{2015MNRAS.452.3209R} to estimate the stellar masses (M$_\star$) of all galaxies in our sample: 
\begin{equation}
  \rm log(M_*/L_*)_g = 2.029 \times (g - r) - 0.984.
\end{equation}  
This relation is based on the \cite{2003MNRAS.344.1000B} stellar population models and the \cite{2003PASP..115..763C} IMF.
Colors are measured from the DECaLS $g$- and $r$-band photometry (see Section~\ref{sec:multi}).
Of the eleven galaxies at $z\sim0.02$, four are dwarfs with M$_*$ $\le$ 10$^9$ M$_{\odot}$ and two -- AGC 983 and IC 0107 -- are massive systems with M$_*$ $\ge$ 10$^{10.5}$ M$_{\odot}$ (Table~\ref{tab:meerkat}, column\,6). The two foreground galaxies at $z\sim0.016$ -- AGC 111931 and AGC 110319 -- are also dwarfs. 

Eight of the ten galaxies with available optical spectra have a median per-pixel (unbinned) S/N $>6$ over 4000$-$9000\,$\rm \AA$. We fit these spectra over 4000$-$9000$\,\rm \AA$ using the Penalized Pixel-Fitting \citep[pPXF;][]{2004PASP..116..138C} software,  with stellar population templates based on the models of \citet{2003MNRAS.344.1000B} and a  \citet{2003PASP..115..763C} IMF. The light-weighted mean metallicities ([M/H]) are also provided in Table~\ref{tab:meerkat}.

The metallicities for the galaxies at $z\sim$0.02 span a wide range, from near-solar values in massive systems (e.g., AGC\,983, [M/H] = 0.02) to metal-poor in dwarfs (e.g., AGC\,111629, [M/H] = $-$1.63). This is consistent with the typical range for galaxies in group environments \citep[][]{2005MNRAS.362...41G, 2020MNRAS.491.5406T}. However, the 3\arcsec\ fiber covers only the central regions, so these values may not represent global metallicities of the galaxies.
Notably, the LSBG AGC\,111629 stands out with an exceptionally low metallicity ([M/H] = $-$1.63), consistent with its diffuse morphology and inefficient star formation history. This aligns with previous findings that LSBGs tend to be more metal-poor than normal galaxies of similar stellar mass \citep{2023ApJ...948...96C, 2024MNRAS.530..812T}.

\subsection{MeerKAT-resolved \HI\ galaxies}
\label{sec:meerkathi}

\setcounter{figure}{3}
\begin{figure*}
  \begin{center}
  \includegraphics[width=7.0in]{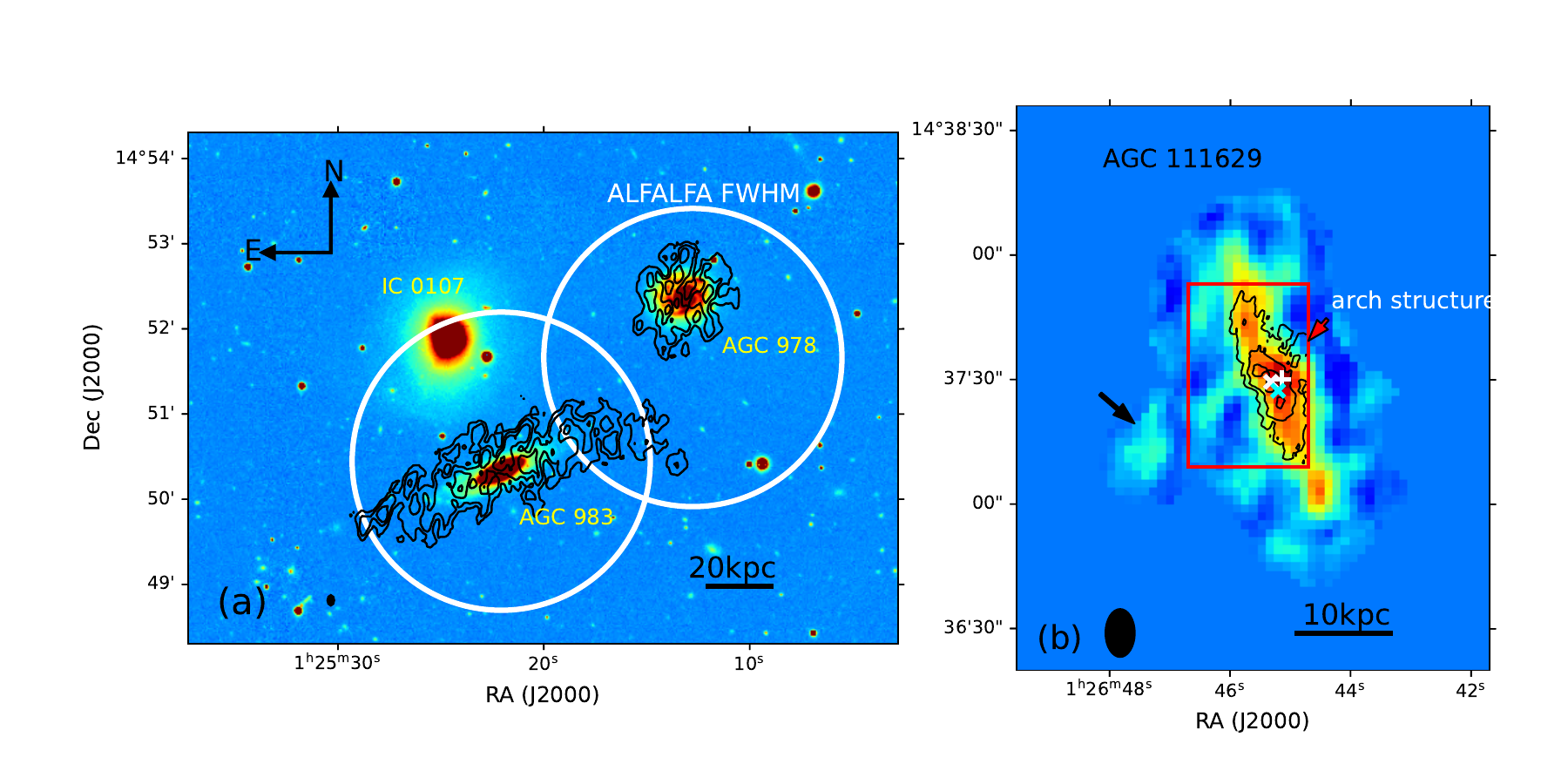}
  \end{center} 
\caption{Panel (a): DECaLS $g$-band image overlaid with \HI\ column density contours from MeerKAT observations of IC\,0107, AGC\,978, and AGC\,983. Contour levels are $ 2^n \times 3.0 \times 10^{20}$ \,cm$^{-2}$ ($n=0,\,2,\,3,\,4,\,5)$. White unfilled circles mark the FWHM of Arecibo telescope (3.5$\arcmin$) used in ALFALFA for AGC\,978 and AGC\,983 \citep{2018ApJ...861...49H}. Panel (b): Integrated H$\alpha$ contours at [3, 12, 36] $\times$ $\sigma$, where $\sigma$ = 8.3 $\times$10$^{-19}$\,erg\,s$^{-1}$\,cm$^{-2}$, overlaid on the integrated \HI, H$\alpha$ and [\ion{O}{III}]$\lambda$5007 peaks, respectively. The red square (29$\arcsec$ $\times$ 44$\arcsec$) shows the H$\alpha$ observation region and the red arrow marks the arch structure observed in H$\alpha$ \citep{2025AJ....170..296C}. The H$\alpha$ spatial resolution is 2.5$\arcsec$ $\times$ 1.1$\arcsec$. The black arrow points to the extended \HI\ structure. The black-filled circle in the lower-left corner of each panel represents the MeerKAT synthesized beam.
\label{fov}}
\end{figure*}

This section presents the properties of the nine galaxies detected by MeerKAT, eight of which have spatially resolved \HI\ maps. Figure~\ref{image1} shows the \HI\ intensity (moment~0), velocity field (moment~1), velocity dispersion (moment~2), optical DECaLS $g$-band, and GALEX NUV images for these eight galaxies. Figure~\ref{spectra} compares the global \HI\ profiles from MeerKAT and ALFALFA for the twelve galaxies detected in at least one survey.
Assuming optically thin emission, the beam-averaged \HI\ column density can be estimated from the \HI\ line intensity using 
\begin{equation}
\label{eqn:hicol}
\frac{N_{\textsc{Hi}}}{\mathrm{cm}^{-2}} = \frac{1.104 \times 10^{24}(1+{\it z})^3}{b_{maj} \times b_{min}} \int \frac{S(v)}{\mathrm{Jy\, beam}^{-1}}\, \frac{dv}{\mathrm{km\,s}^{-1}},
\end{equation}
where $b_{maj}$ and $b_{min}$ are major and minor axes of the restoring beam in arcseconds, with $v$ being measured in the reference frame of the target.
For a spectral rms of 0.8\,mJy\,beam$^{-1}$ per 5.5\,\kms\ channel and primary beam attenuation correction in the range 0.4 -- 0.8, this yields a 1$\sigma$ column density sensitivity of 1 -- 2 $\times 10^{20}$\,cm$^{-2}$.

For non-detected targets, the upper limits on the \HI\ flux are estimated as following. The non-detection \HI\ spectra are smoothed to a velocity resolution ($\Delta v$) of 20\,\kms, and the rms noise, $\sigma$, is measured within a 300\,\kms\ velocity window centered on the target. The 3$\sigma$ upper limit on the integrated \HI\ flux is then estimated as:
\begin{equation}
\label{eqn:hifluxlimit}
S_{3\sigma} = 3 \times \sigma \times \Delta v \times \sqrt{N},
\end{equation}
where N is W$_{50}$/$\Delta v$. For AGC\,112961, AGC\,110339, and AGC\,748815, the W$_{50}$ values in Table~\ref{tab:meerkat} are based on FASHI measurements (private communication). The W$_{50}$ for IC\,0107 is assumed 20\,\kms.
The \HI\ masses are calculated using:
\begin{equation}
\label{eqn:himass}
\frac{M_{\textsc{Hi}}}{M_\odot} = \frac{2.356 \times 10^5}{1 + z} \times\bigg({\frac{D_L}{\rm{Mpc}}}\bigg)^2\, \int{\frac{S(v)}{\mathrm{Jy}}\,\frac{dv}{\mathrm{km\,s}^{-1}}},
\end{equation}
\noindent
where $D_L$ is the cosmological luminosity distance corresponding to the redshift of the galaxy.
The \HI\ properties of these galaxies derived from the MeerKAT data are provided in Table~\ref{tab:meerkat}.  
The M$_{\textsc{Hi}}$ uncertainties were propagated from the integrated-flux and distance uncertainties following Equation~(5) of \citet{2018ApJ...861...49H}, with a 5\% absolute flux-calibration uncertainty \citep{2021ApJ...907...11G} added in quadrature.
%

For seven of the eight MeerKAT-resolved galaxies, the \HI\ fluxes derived from MeerKAT and ALFALFA agree to within $\sim$20\%, which is consistent with the level of agreement generally found in direct MeerKAT–ALFALFA comparisons \citep[e.g.,][]{2021A&A...654A.173H}.
The exception is AGC\,978, for which the ALFALFA beam encompasses both AGC\,978 and the western disk of AGC\,983, leading to an overestimate of the AGC\,978 flux in ALFALFA and a corresponding underestimate for AGC\,983, as discussed further in Section~\ref{sec:group}.
%

We also derive SFRs using the 1.4\,GHz radio continuum images from MALS data release 1 \citep[DR1;][]{Deka24dr1}. The SFR calibration follows \cite{2011ApJ...737...67M}, adopting 6.35 $\times$ 10$^{-29}$ L$_{1.4GHz}$ (erg s$^{-1}$ Hz$^{-1}$) assuming \cite{2001MNRAS.322..231K} IMF, and the resulting values are converted to a \cite{2003PASP..115..763C} IMF by applying SFR$^{Cha}$ = SFR$^{Kro}$ $\times$ 0.943 \citep{2017A&A...599A..71D}, as presented in Table~\ref{tab:meerkat}. 
Five of the eight MeerKAT-resolved galaxies have radio-continuum detections.
For three of them, namely AGC\,978, AGC\,1027 and AGC\,111930, the radio-derived SFRs agree to within a factor of two with those inferred from the dust-corrected NUV method, consistent with the typical scatter between the two independent tracers.
The two exceptions, AGC\,983 and AGC\,111931, can be understood in terms of the different timescales probed by the two SFR tracers.

In the following, we briefly describe the individual properties of the group members ($z\sim 0.02$; Section~\ref{sec:group}) and the two foreground galaxies ($z\sim 0.016$; Section~\ref{sec:foreground}). The \HI\ disk geometry and inclinations are derived in Section~\ref{sec:geometry}, the resolved kinematic modelling is presented in Section~\ref{sec:modle}, and the radial \HI\ surface-density profiles are discussed in Section~\ref{sec:hi_profiles}.

\subsubsection{Group members ($z\sim0.02$)}
\label{sec:group}

\noindent   \textbf{AGC\,983 and AGC\,978:}
AGC\,983, also known as IC\,1698, is a ring-barred galaxy \citep{2020MNRAS.491.3767S}, hosting an Active Galaxy Nucleus (AGN)  identified via the BPT diagram \citep{2019ApJ...872..134Z}. It exhibits a red color and solar metallicity in its central region (Table~\ref{tab:meerkat}). AGC\,112961 is at a similar velocity to AGC\,983 (Figure~\ref{DSP}), with a projected separation of 357\,kpc. The W${50}$ of AGC\,983 is 431.4\,\kms, and its western disk reaches velocities of $\sim$6400\,\kms (see moment-1 map in Figure~\ref{image1}). This is very close to the systemic velocity of AGC\,978. The radio-derived SFR of AGC\,983 exceeds the dust-corrected NUV-based value by about a factor of 3.4, likely due to additional non-thermal emission from its central AGN boosting the 1.4\,GHz continuum.

AGC\,978 and IC\,0107 are at the same systemic velocity. IC\,0107 is the most massive galaxy in Figure~\ref{DSP} and hosts an AGN \citep{2019ApJ...872..148C}. It is not detected in \HI\ by MeerKAT and ALFALFA observations. The \HI\ spectrum of AGC\,978 from MeerKAT shows a clear double-horned profile, differing from the ALFALFA spectrum. The MeerKAT and ALFALFA integrated \HI\ fluxes differ by 30\%. As shown in Figure~\ref{fov}(a), the ALFALFA beam encompasses both AGC\,978 and the western disk of AGC\,983 at $\sim$6400\,\kms, leading to flux contamination in the single-dish spectrum.

\noindent    \textbf{AGC\,1027:}
AGC\,1027 (IC\,1704) is a galaxy with clear spiral arms in the optical image. Its nearest neighbor at a similar velocity is AGC\,110339, at a projected distance of 153\,kpc. The clumpy structure along the spiral arms corresponds to star-forming regions. AGC\,1027 has colors, stellar mass, and central [M/H] very similar to AGC\,978. In both galaxies, the \HI\ surface density is highest in the inner regions and declines outward (see Section~\ref{sec:hi_profiles}). 

\noindent    \textbf{AGC\,111629:}
AGC\,111629 is a LSBG with an irregular H$\alpha$ morphology and an arch-like structure in the western extra-planar region \citep{2025AJ....170..296C}. The irregular H$\alpha$ morphology may be related to a past merger with its satellite galaxy AGC\,748815, located to the northeast, with a diffuse structure visible between them. A low-luminosity AGN may exist at the center of AGC\,111629 \citep{2025AJ....170..296C}. 
Figure~\ref{fov}(b) shows the H$\alpha$ contours overlaid on the integrated \HI\ image of AGC\,111629. The \HI\ emission is broadly consistent with the H$\alpha$ morphology. The \HI\ peak (white plus marker) lies close to the H$\alpha$
and [\ion{O}{III}]$\lambda$5007 peaks \citep{2025AJ....170..296C}, with a slight offset that may be due to the coarser resolution of the radio data. Near the H$\alpha$ arch in AGC\,111629, northwest of the \HI\ peak, the native-resolution moment-2 map shows a local line-of-sight velocity dispersion of up to about $80$\,\kms\ (Figure~\ref{image1}). The $15''$ \textsc{3DBarolo} model measures the intrinsic dispersion on the scale of a ring, yielding 11.9 and 11.4\,\kms\ in the inner and outer rings, respectively; the high moment-2 values likely arise from noise in this low-S/N region or from multiple velocity components.

Notably, an extended \HI\ structure stretching toward the southeast is clearly detected without a bright optical counterpart (DECaLS $g$-band image in Figure~\ref{image1}). This asymmetric extension, highlighted by the black arrow in Figure~\ref{fov}(b), may be consistent with a tidal feature or a remnant of the merger event discussed in \cite{2025AJ....170..296C}.

\noindent \textbf{AGC\,111930 and AGC\,112965:}
AGC\,111930 is classified as an AGN by the MaNGA survey \citep{2015ApJ...798....7B, 2023A&A...674A..85A}. Its optical and the resolved \HI\ morphology show an irregular structure. It has a satellite galaxy, AGC\,112965, at a projected distance of 198\,kpc, which is classified as a starburst galaxy by SDSS fiber spectrum. 
As with AGC\,111629 and its satellite, the star formation activity in AGC\,112965 may have been triggered by tidal interaction with AGC\,111930. The irregular \HI\ morphology of AGC\,111930 may also be a consequence of this interaction.

\noindent \textbf{AGC\,748816:}
AGC\,748816 is an \HI-bearing UDG identified by \citet[][]{2017ApJ...842..133L} and \citet[][]{2019MNRAS.490..566J}. The galaxy shows a discernible stellar component in the central region, together with an identifiable outer disk structure. It has a very blue color ($g − r$ = $-$0.04) and low stellar mass (M$_\star$ = 3.9$\times$ 10$^7$ M$_{\odot}$), consistent with its classification as a gas-rich, star-formation-deficient dwarf. The resolved \HI\ morphology shows a regular, roughly circular distribution with no obvious signs of tidal disturbance. 
The global \HI\ profile of AGC\,748816 is single-peaked. Such a profile may result from a relatively low inclination, a substantial contribution from velocity dispersion, or a slowly rising rotation curve. However, a kinematically cold rotating disc can also appear single-peaked when its internal rotational structure is insufficiently resolved. Given the limited spatial resolution of the present observations, the profile shape alone cannot distinguish among these possibilities. The inclination and rotation velocity are constrained independently in Sections~\ref{sec:geometry} and~\ref{sec:modle}.
The \HI\ mass is significantly higher than expected for its stellar mass, placing it well above the M$_\star$-M$_{\textsc{Hi}}$ relation in Figure~\ref{relation}(b).

\subsubsection{Foreground galaxies ($z\sim0.016$)}
\label{sec:foreground}

AGC\,111931 and UDG AGC\,110319 are at the same velocity, separated by a projected distance of $\sim$314\,kpc. 
AGC\,111931 is an edge-on galaxy with irregular \HI\ and optical morphologies in its outer disk. Its radio-based SFR is significantly lower than the dust-corrected NUV SFR. 
 We note that the radio continuum at 1.4\,GHz traces synchrotron emission from cosmic-ray electrons accelerated in supernova remnants, providing a measure of star formation over timescales of $\sim$100-300\,Myr \citep[][]{1992ARA&A..30..575C, 2023A&A...675A.12A}, comparable to or longer than the NUV, which is sensitive to star formation over $\sim$100\,Myr \citep[][]{2012ARA&A..50..531K}.
In low-mass galaxies, cosmic-ray electrons can escape the galaxy before losing energy to synchrotron radiation, suppressing the 1.4\,GHz luminosity relative to the true SFR \citep[e.g.,][]{2008ApJ...678..828M}. Alternatively, this discrepancy can be explained by a recent star formation episode with which the radio synchrotron emission has not yet fully caught up.

UDG AGC\,110319 has a redder color than AGC\,111931 and a central \HI\ deficit. 
Its dust-corrected H$\alpha$- and NUV-based SFRs are 0.034\,M$_{\rm \odot}$\,yr$^{-1}$ \citep{2018ApJS..235...18L} and 0.087\,M$_{\rm \odot}$\,yr$^{-1}$, respectively. Optical imaging from the DESI Legacy Survey reveals that UDG AGC\,110319 is viewed at a low inclination angle. The central region contains a prominent oval-shaped stellar structure, while the outer stellar disk is difficult to trace.
We note that the apparent \HI\ deficit may be partly due to the limited depth of the current data; nevertheless, the central region still exhibits a significantly lower \HI\ content.
The irregular \HI\ morphologies of both foreground galaxies are also suggestive of a past interaction. More sensitive \HI\ imaging is needed to further investigate this scenario.

\subsection{\HI\ disk geometry}
\label{sec:geometry}

A reliable measurement of disk inclination is essential both for deprojecting the observed \HI\ morphology and for correcting rotation velocities in the kinematic analysis presented in Section~\ref{sec:modle}. We derive inclinations from the resolved \HI\ moment-0 distribution at native resolution, described below, and compare these with optical-based estimates later in this section. We measure the inclination of the regular main \HI\ body while minimizing the influence of detached clouds, asymmetric low-surface-brightness extensions, central depressions, and compact high-column-density clumps. Because a single ellipse fitted to the full \HI\ extent can be biased by such structures, we adopt a beam-corrected multi-isophote method based only on contours that consistently trace the same main \HI\ structure.

The moment-0 maps were mildly smoothed with a circular Gaussian kernel. The smoothing width was allowed to vary uniformly between 0.9 and 1.1 image pixels to test the dependence of the fitted geometry on the smoothing scale. We then scanned a broad range of isophotal levels, expressed as fractions of the peak value of the smoothed moment-0 map. At each level, we selected the contour associated with the main \HI\ body using the same criteria for all galaxies: a fitted center within ten pixels of the reference \HI\ center (Table~\ref{tab:additional_data}), and an enclosed ellipse area of at least 13 synthesized beams. This area threshold corresponds to  $\sim$4.3 beam FWHMs, and was chosen to limit sensitivity to individual internal clumps and beam-correction uncertainties. Detached contours, and contours dominated by a single internal clump, were rejected.

We defined a stable interval as a contiguous set of isophotes that traced the same connected \HI\ structure, showed no abrupt change in fitted center, and produced a smoothly varying beam-corrected axis ratio. Isophotes that became connected to an outer extension, switched to another internal component, or showed a sudden
center displacement were excluded. The adopted peak-fraction intervals ranged from 18--31\% for most galaxies: AGC\,983 (19--26\%), AGC\,978 (20--27\%), AGC\,1027 (18--26\%), AGC\,748816 (19--26\%), AGC\,111930 (19--23\%), AGC\,111931 (20--31\%), and 5--10\% for AGC\,111629 (LSBG), whose markedly narrower and lower-level interval reflects its more irregular \HI\ morphology (Section~\ref{sec:group}). AGC\,110319 (UDG), the eighth \HI-resolved galaxy, is treated separately below. Although the numerical intervals differ, the same center, minimum-area, and stability criteria were used throughout.

An ellipse was fitted to each accepted contour and corrected for the synthesized beam and the additional smoothing kernel using an equivalent two-dimensional covariance-matrix representation. The intrinsic projected axis ratio was converted into an inclination by assuming an oblate disk with an intrinsic thickness of $\rm q_0$ = 0.2 \citep[e.g.,][]{2009AJ....138..323T, 2020ApJ...899...78R}. The full beam-deconvolution procedure and the corresponding equations are described in Appendix~\ref{app:hi_geometry_method}. The resulting inclination ($i_{\textsc{HI}}$; column\,14 of Table~\ref{tab:meerkat}) is a morphological inclination of the regular main \HI\ disk rather than an independent kinematic inclination derived from the velocity field.

\begin{figure*}[ht]
  \begin{center}
  \includegraphics[width=7.0in]{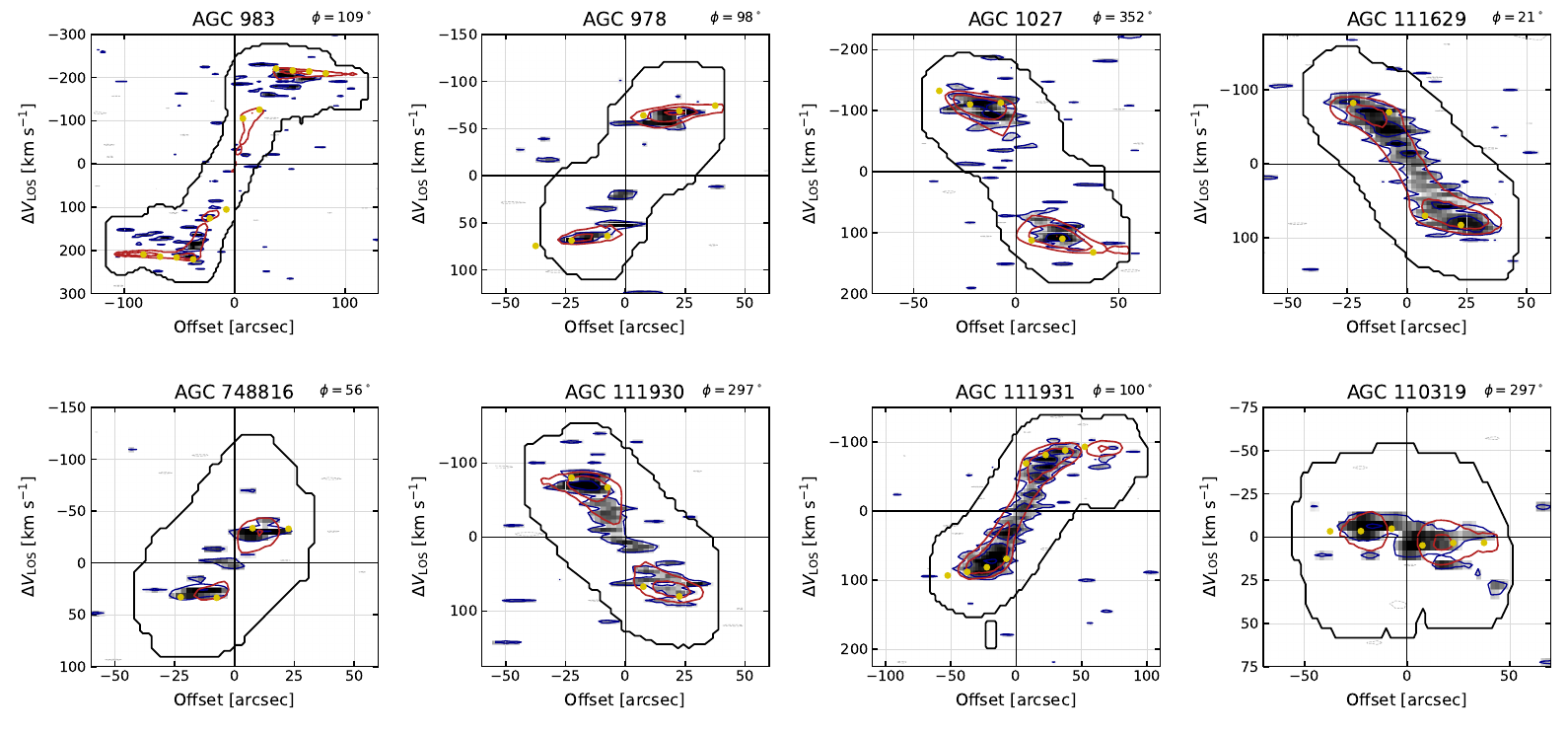}
  \end{center} 
  \caption{Major-axis position--velocity diagrams for the eight galaxies
in the MeerKAT \HI\ cubes smoothed to a circular beam of $15''$ FWHM.
The panels follow the order of Table~\ref{tab:meerkat}. Velocities are
shown relative to the systemic velocity of each galaxy. Grayscale and
blue contours show the observed \HI\ emission along the adopted major
axis, and red contours show the corresponding \textsc{3DBarolo} model.
The positive contour levels are $2.5 \times 2^n \sigma_{\rm cube}$
for both data and model, where $n=0,1,2,\ldots$ and $\sigma_{\rm cube}$
is the rms noise per channel in the observed cube. The faint dashed
gray contour marks $-2.5\sigma_{\rm cube}$ in the data. Yellow points
show the projected rotation velocities of the fitted rings, and the
black contour outlines the fitting mask. The adopted major-axis PA
is given in the upper right corner of each panel. For AGC\,110319,
the rotation signal is too weak to yield a reliable circular velocity.}
\label{3DBarolo}
\end{figure*}

Internal numerical uncertainties were assessed by repeating the full fitting procedure for 10,000 independent realizations, randomly perturbing the isophotal level, smoothing scale, and contour coordinates at each step. The resulting inclinations were stable, with formal scatter generally below 1$^{\circ}$. This internal check, however, does not capture systematic effects from limited spatial resolution, non-axisymmetric \HI\ structure, intrinsic disk thickness, or the adopted geometrical model. We therefore adopt a conservative inclination-uncertainty floor of 5$^\circ$ for galaxies with regular \HI\ morphology, motivated by the mock-observation tests of \citet{2019ApJ...883L..33M}, who found that inclinations could be recovered to within approximately 5$^\circ$ for $i \gtrsim$ 30$^\circ$ in \HI-rich dwarf galaxies observed at comparable spatial resolution and S/N. This value is an adopted systematic floor, not a formal confidence interval from the Monte Carlo realizations. AGC\,748816, with  $i_{\textsc{HI}}$ = 29$^\circ$, lies near the lower-inclination limit of these tests. We retain the same 5$^\circ$ floor for this galaxy, noting that its inclination-dependent uncertainties may be somewhat larger in practice.

As an independent consistency check, we compared the \HI-based inclinations with the optical inclinations, which were derived from \textsc{SExtractor} axis ratios measured in the DECaLS $g$-band images and the same $\rm q_0$ = 0.2. The two estimates agree for six of the eight \HI-resolved galaxies. The exceptions are AGC\,111629 (LSBG) and AGC\,110319 (UDG), whose irregular and non-axisymmetric optical morphologies make a single $g$-band axis ratio an unreliable tracer of the outer disk geometry. In particular, the measurement for AGC\,110319 is strongly influenced by its central oval stellar structure, while its outer stellar disk is too faint to be robustly traced.

We now consider AGC\,110319, which was treated separately because its moment-0 map does not show a unique sequence of isophotes that can be unambiguously associated with a regular main \HI\ disk. Different contour levels trace the central depression and the surrounding clumpy emission causing the fitted centre and axis ratio to depend strongly on the selected isophotal range. We therefore retained the inclination obtained from a simplified forward-modeling approach similar to \textsc{SOCCA} \citep{DiMascolo2026}, in which the full masked moment-0 distribution was fitted with a single inclined, axisymmetric disk model convolved with the synthesized beam. The disk centre, position angle, inclination, and radial surface-brightness scale were allowed to vary. Because this global model cannot fully describe the central depression and clumpy, non-axisymmetric emission, the inclination posterior is broad, with a 1$\sigma$ uncertainty of $\sim14^\circ$. We retain this larger uncertainty as a more realistic representation of the geometrical ambiguity in AGC\,110319. The adopted $i_{\textsc{HI}}$ are listed in Table~\ref{tab:meerkat} (column\,14).

\subsection{Three-dimensional \HI\ kinematic modeling}
\label{sec:modle}

We used \textsc{3DBarolo} to fit tilted-ring models to the resolved \HI\ emission. To assess how the results depend on resolution and signal-to-noise ratio, we fitted three versions of the data: the native-resolution cubes ($12\farcs1\times7\farcs2$, 5.5\,\kms\ channels), cubes smoothed over three adjacent channels ($\sim$16.5\,\kms) at the native beam, and cubes smoothed to a circular beam of $15''$ FWHM. Spectral smoothing increased the S/N per channel but did not systematically improve the agreement between the observed and model position--velocity diagrams or channel maps. Spatial smoothing to $15''$ produced more continuous emission while retaining $\sim$5--16 beams across the major axes, and yielded a stable fit for AGC\,111930, which had none at native resolution. We therefore adopted the $15''$ cubes for the final fits. For six of the seven galaxies with measurable rotation, the outermost fitted rotation speeds differ by at most 4.6\,\kms\ between the native and $15''$ cubes. For AGC\,1027, $V_{\rm rot}$ at $R=37.5''$ increases from 179 to 206\,\kms. Refitting the $15''$ cube with the native-resolution mask gives 177\,\kms, indicating that the outer-ring speed is sensitive to the emission included in the fit. The rotation speed of AGC\,110319 remains poorly constrained in all three fits.
Results that remain sensitive to the modelling choices, namely the two-ring galaxies, the outermost ring of AGC\,978, and the adopted dispersions, are discussed below. The smoothed cubes were used only for the kinematic modelling; detections and the \HI\ properties in Table~\ref{tab:meerkat} are based on the native-resolution cube. We adopted the disk centers measured from the native-resolution moment-0 maps and the inclinations derived in Section~\ref{sec:geometry}.

We determined the position angles (PAs) at native resolution. 
For AGC\,983 and AGC\,111930, the PA fitted freely in each ring was consistent across adjacent rings and different starting values. We adopted the median of these values. For AGC\,978, AGC\,1027, AGC\,111629, and AGC\,111931, we measured the direction from the weighted centroid of the approaching side to that of the receding side in the moment-1 map. For AGC\,748816, the free-PA fits were sensitive to the starting value. We adopted $56^\circ$, consistent with the gradient direction of $56.3^\circ$ from a planar fit to the native-resolution velocity field. For AGC\,110319, the velocity gradient was weak. We measured the morphological major axis from the global disk model of Section~\ref{sec:geometry} and used the velocity field to identify the receding side, giving an adopted PA of $297^\circ$. The adopted PAs are listed in column~15 of Table~\ref{tab:meerkat}.

We measured the systemic velocity, $V_{\rm sys}$, from each galaxy's global MeerKAT \HI\ profile shown in Figure~\ref{spectra}. We smoothed each profile with a Gaussian kernel of standard deviation one spectral channel. On each outer flank, we linearly interpolated between adjacent channels to find the velocity at 50\% of the peak flux density on that side. For the single-peaked profile of AGC\,110319, we used the same peak for both flanks. We adopted the midpoint of the two crossing velocities as $V_{\rm sys}$, which is listed in column~10 of Table~\ref{tab:additional_data}.

In the final fits, we held the disk center, inclination, PA, and systemic velocity fixed at all radii. Only the rotation speed, $V_{\rm rot}$, and the \HI\ velocity dispersion, $\sigma_{\rm HI}$, were free to vary independently from ring to ring. We set the radial velocity and vertical scale height to zero, restricted the fits to the three-dimensional source masks, and used local normalization (\texttt{NORM=LOCAL}) to scale the model flux to the observed emission. Adjacent rings were separated by $15''$, one beam FWHM. We fitted two rings for AGC\,111629, AGC\,748816, and AGC\,111930, three for AGC\,978, AGC\,1027, and AGC\,110319, four for AGC\,111931, and six for AGC\,983. Figure~\ref{3DBarolo} compares the observed and modeled major-axis position--velocity diagrams for all eight galaxies at $15''$ resolution. The models broadly trace the observed major-axis velocity gradients in seven of the eight galaxies. AGC\,111629 and AGC\,111931 show the clearest continuous emission across the systemic velocity. The emission is patchier in AGC\,983, AGC\,978, AGC\,1027, and AGC\,111930. In AGC\,748816, the emission is faint, but the approaching and receding sides are distinguishable. For AGC\,110319, the projected velocity gradient is too weak to constrain the rotation speed reliably.

For AGC\,983, AGC\,978, AGC\,1027, and AGC\,111931, we derived circular speeds, $V_{\rm circ}$, using the asymmetric-drift correction in \textsc{3DBarolo}, which accounts for pressure support using the radial profiles of \HI\ surface density and velocity dispersion \citep{2017MNRAS.466.4159I}. 
For all galaxies, we report the value at the outermost ring where the observed major-axis position--velocity diagram shows emission at or above $2.5$ times the cube rms on both sides near the projected model velocities (hereafter the bilateral emission criterion). For the three galaxies fitted with only two rings, the radial gradients required for this correction were poorly constrained. We therefore report the fitted outer-ring $V_{\rm rot}$ for AGC\,111629, AGC\,748816, and AGC\,111930. Column~16 of Table~\ref{tab:meerkat} gives the adopted speeds. 
For each speed, we combined the ring-fit uncertainty in quadrature with the change in speed obtained by refitting at inclinations of $i\pm5^\circ$, evaluating the upper and lower uncertainties separately.

The fitted dispersions in the outermost rings range from 1.0 to 11.4\,\kms. The values for AGC\,983 and AGC\,978 lie at the lower fitting limit of 1\,\kms\ and are poorly constrained. To test how these low dispersions affect their circular speeds, we repeated the fits with $\sigma_{\rm HI}$
fixed in all rings at 8 or 10\,\kms. In each case, we refitted $V_{\rm rot}$ and recomputed the asymmetric-drift correction.  
At the radii at which the velocities in Table~\ref{tab:meerkat} are measured ($52.5''$ for AGC\,983 and $22.5\arcsec$ for AGC\,978), the fixed-dispersion fits give $V_{\rm circ}=223.2$--$223.3$\,\kms\ and $103.0$--$103.1$\,\kms, respectively.
These values differ from the free-dispersion results by at most 3.1\,\kms\ for AGC\,983 and 0.3\,\kms\ for AGC\,978, both within the quoted uncertainties. The outermost fitted ring of AGC\,978 ($37.5''$) was more sensitive to the assumed dispersion: $V_{\rm circ}$ decreased from 111.3\,\kms\ in the free-dispersion fit to 74.4 and 71.0\,\kms\ in the two tests. This ring did not meet the bilateral emission criterion used for Table~\ref{tab:meerkat}.

AGC\,110319 has a narrow global \HI\ profile, with $W_{50}=20.8$\,\kms\ (Table~\ref{tab:meerkat}). The fitted rotation speeds of 6.5--9.4\,\kms\ are comparable to or smaller than the fitted gas dispersion in each of the three rings. At $i=31^\circ$, the projected speeds, $V_{\rm rot}\sin i$, are only 3.3--4.8\,\kms, below the channel spacing of 5.6\kms\ at this redshift. 
Given these small projected speeds and the contribution of random motions to the global line width, neither the tilted-ring fit nor the line width provides a reliable rotation speed for this galaxy. 
We therefore exclude AGC\,110319 from velocity-based analyses.

\subsection{Radial \HI\ surface-density profiles}
\label{sec:hi_profiles}

\begin{figure*}[ht]
  \begin{center}
  \includegraphics[width=7.0in]{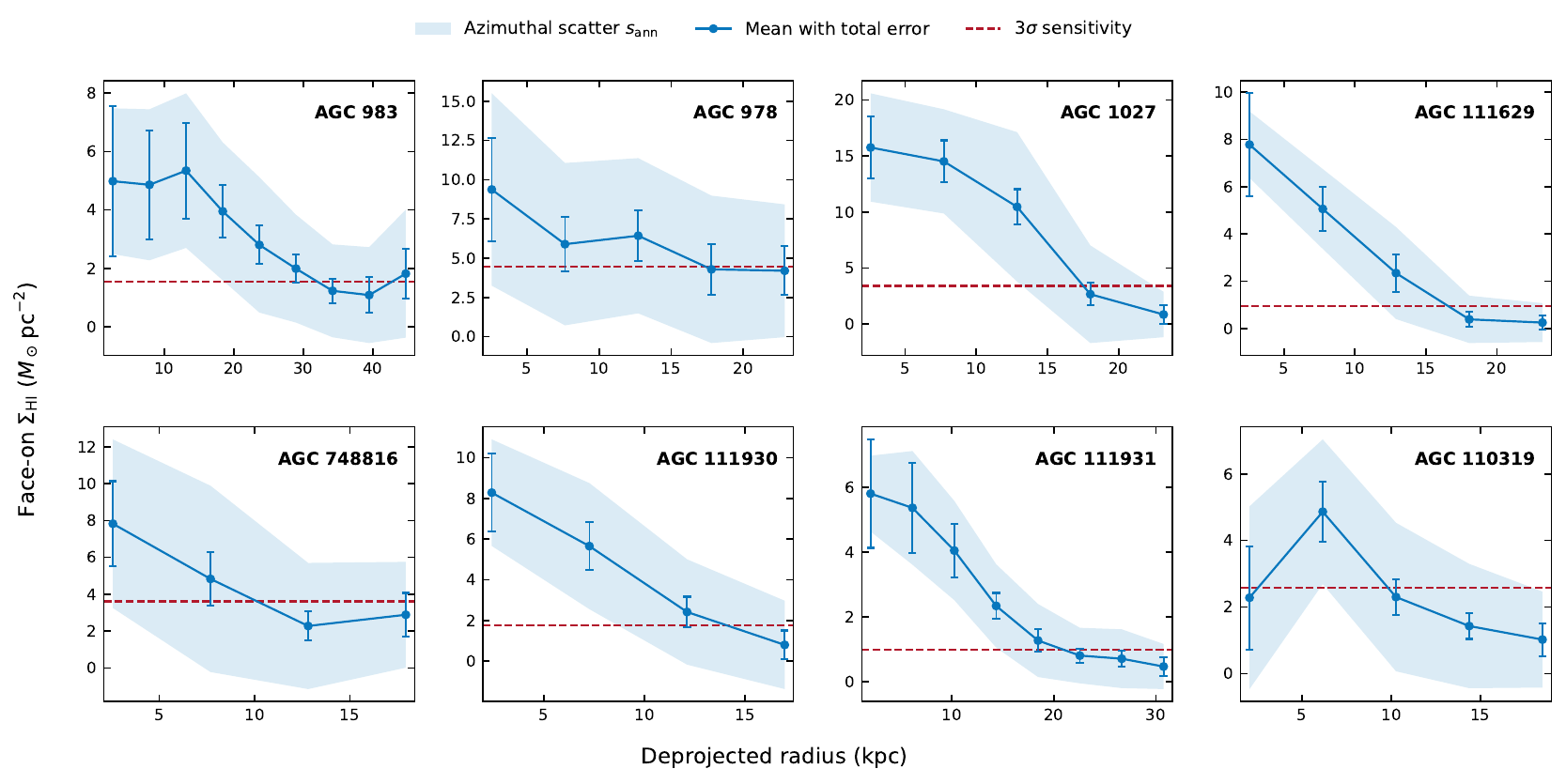}
  \end{center} 
  \caption{Azimuthally averaged, face-on \HI\ surface-density profiles of the eight galaxies in our sample. The profiles are sampled using deprojected elliptical annuli with a radial width equal to one synthesized-beam major-axis FWHM of $12\farcs1$.} Blue points show the mean \HI\ surface density within each deprojected elliptical annulus, while the error bars combine the uncertainty on the annular mean and that propagated from the adopted inclination uncertainty. Light-blue shaded regions show the azimuthal standard deviation within each annulus, tracing both measurement noise and genuine structure such as \HI\ clumps, holes, and large-scale disc asymmetries. The horizontal dashed line in each panel marks the galaxy-specific 3$\sigma$ face-on surface-density sensitivity, accounting for the primary-beam attenuation and inclination at that galaxy's position.
\label{surface}
\end{figure*}

We derived azimuthally averaged \HI\ surface-density profiles from the moment-0 maps, adopting the same fixed centres used for the \textsc{3DBarolo} modelling where available. AGC\,110319 required separate treatment: its nearly face-on clumpy \HI\ distribution with a central depression makes neither the brightest \HI\ pixel nor the intensity minimum a robust estimate of the disk centre. 
For the radial profile of AGC\,110319 in Figure~\ref{surface}, we centred the annuli on the SoFiA \HI\ source position ($\mathrm{RA}=01^{\rm h}25^{\rm m}16.62^{\rm s}$, $\mathrm{Dec}=+14^\circ08'53.7''$) and adopted a PA of $297^\circ$. This centre lies $2.1''$ from the centre adopted for the kinematic fit, which we derive from the global disk model of Section~\ref{sec:geometry}. The separation is smaller than the $7.2''$ minor axis of the native beam.

Using the adopted centre, position angle, and inclination, we assigned each image pixel a deprojected galactocentric radius following the standard tilted-ring geometry \citep{Begeman1989,DiTeodoro2015,Oh2018}. The face-on \HI\ surface density was calculated from the velocity-integrated 21-cm intensity assuming optically thin emission \citep{Walter2008,Leroy2008}. The profiles were then measured in concentric deprojected elliptical annuli with a radial width equal to the synthesized-beam major-axis FWHM $12\farcs1$, comparable to the ring separations of the kinematic models (Section~\ref{sec:modle}), and providing approximately independent radial measurements along the major axis. Negative pixels were retained to avoid a positive bias, and annuli with less than 80 per cent image coverage were excluded. The detailed deprojection and surface-density conversion are provided in Appendix~\ref{app:hi_profiles}.

For each annulus, we quantified the azimuthal scatter as the standard deviation of the \HI\ surface-density values. The statistical uncertainty on the annular mean was then estimated by dividing this scatter by the square root of the number of independent synthesized beams within the annulus. Because the scatter reflects both measurement noise and real non-axisymmetric structure, such as \HI\ clumps, local depressions or holes, and large-scale disc asymmetries, this estimate should not be interpreted as a pure thermal-noise uncertainty. We propagated the inclination uncertainty by repeating the entire profile extraction at the lower and upper inclination limits. The resulting inclination-induced uncertainty was then added in quadrature to the uncertainty on the annular mean to derive the error bars shown in the radial profiles (Figure~\ref{surface}).

The \HI\ surface-density profiles of nearby gas-rich galaxies typically have central values of a few to approximately 10\,M$_\odot\,{\rm pc^{-2}}$ and decline toward larger radii, becoming approximately self-similar when scaled by a characteristic \HI\ radius \citep{2016MNRAS.460.2143W,2025ApJ...980...25W}.
Most galaxies in our sample broadly follow this behavior, although their resolved peak surface densities span a relatively wide range of approximately 5--20\,M$_\odot\,{\rm pc^{-2}}$ and local variations are present because of clumpy or asymmetric gas distributions (Figure~\ref{surface}). The homogeneously sampled, beam-matched profiles of the two UDGs show different radial structures. AGC\,748816 has its highest \HI\ surface density in the central annulus, $\Sigma_{\rm HI}\simeq$ 7.8\,M$_\odot\,{\rm pc^{-2}}$, followed by an overall decline toward larger radii. In contrast, AGC\,110319 has $\Sigma_{\rm HI}\simeq$ 2.4\,M$_\odot\,{\rm pc^{-2}}$ in the central
annulus and reaches a maximum of approximately 4.8\,M$_\odot\,{\rm pc^{-2}}$ in the second annulus, at a deprojected radius of approximately 6\,kpc. This profile is consistent with an apparent central \HI\ depression and with the clumpy, non-axisymmetric morphology seen in its moment-0 map. However, the difference between the innermost bins is not statistically significant given the uncertainties, and the central value is close to the corresponding 3$\sigma$ sensitivity. The central depression should therefore be regarded as tentative. The two UDGs nevertheless illustrate the diversity of atomic-gas distributions among such systems rather than a single characteristic radial \HI\ profile.

\section{Discussion}
\label{sec:disc}

\begin{figure*}
  \begin{center}
  \includegraphics[width=\textwidth]{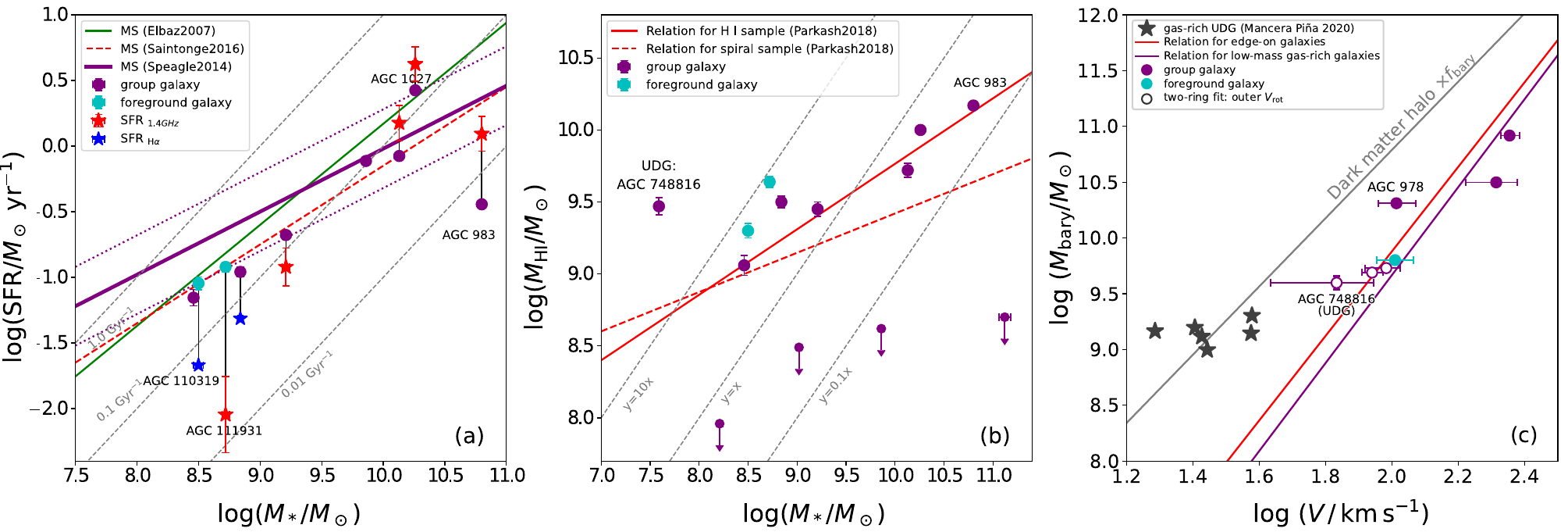}
  \end{center} 
\caption{Panel (a): The M$_*$-SFR relation. Colored circles show SFR derived from the dust corrected NUV luminosity for group and foreground galaxies from Table~\ref{tab:meerkat}. Red stars represent the SFR derived from 1.4\,GHz continuum (Table~\ref{tab:meerkat}). Blue stars represent the SFRs derived from H$\alpha$ from \cite{2018ApJS..235...18L} and \cite{2025AJ....170..296C}.
The purple, green and red lines show the star-forming main sequence from \cite{2007A&A...468...33E}, \cite{2014ApJS..214...15S} and \cite{2016MNRAS.462.1749S}, respectively. These provide a representative range of the expected star formation activity for galaxies across different environments and redshifts. The gray dashed lines, arranged from bottom to top, correspond to specific star formation rates (sSFRs) of 0.01\,Gyr$^{-1}$, 0.1\,Gyr$^{-1}$, and 1\,Gyr$^{-1}$. 
All SFRs were converted to a \cite{2003PASP..115..763C} IMF using SFR$^{Cha}$ = SFR$^{Sal}$ / 1.41 \citep{2007A&A...468...33E} and SFR$^{Cha}$ = SFR$^{Kro}$ $\times$ 0.943 \citep{2017A&A...599A..71D}. 
Panel (b): The M$_*$-M$_{\textsc{Hi}}$ relation. Colored circles show M$_*$ and M$_{\textsc{Hi}}$ for group and foreground galaxies from Table~\ref{tab:meerkat}. The red solid and dashed lines show the relations followed by the \HI-selected sample of galaxies and the spiral galaxies from the M$_*$-selected sample, respectively, each with 1$\sigma$ scatters of 0.5\,dex and 0.4\,dex \citep{2018ApJ...864...40P}.
Panel (c): Baryonic Tully-Fisher relation. Baryonic mass as a function of velocity for the seven
galaxies with velocity measurements in column~16 of Table~\ref{tab:meerkat}. Purple and cyan circles denote group and foreground galaxies, respectively. Filled circles show the asymmetric-drift-corrected circular speeds ($V_{\rm circ}$) of AGC\,983, AGC\,978, AGC\,1027, and AGC\,111931. Open circles show the fitted outer-ring rotation speeds ($V_{\rm rot}$) of AGC\,111629, AGC\,748816, and AGC\,111930, for which the two rings
in the $15''$ models provide insufficient radial information to constrain the asymmetric-drift correction. AGC\,110319 is omitted because its rotation speed is poorly constrained. Black stars denote the gas-rich
UDGs from \citet{2020MNRAS.495.3636M}. The red and purple lines show the best-fitting relations for edge-on ALFALFA galaxies \citep{2016A&A...593A..39P} and low-mass gas-rich galaxies \citep{2009A&A...505..577T}, respectively. The gray line shows $M_{\rm bary}=f_{\rm bary}M_{200}$ for the cosmic baryon fraction
$f_{\rm bary}=0.157$ \citep{2012AJ....143...40M}.}
\label{relation}
\end{figure*}

\subsection{Star formation and \HI\ gas content}

We show the M$_*$-SFR relation and M$_*$-M$_{\textsc{Hi}}$ relation in Figure~\ref{relation} (a) and (b). The SFRs of our sample are mainly near and below the main sequence (MS), with the exception of AGC\,1027, which is actively star-forming. AGC\,983 stands out as the most strongly suppressed, with a specific star formation rate ($\sim$0.006\,Gyr$^{-1}$ based on NUV SFR) roughly an order of magnitude below the rest of the sample, although its total \HI\ content is close to that expected for its stellar mass. Its relatively low \HI\ surface density suggests that much of the atomic gas is distributed diffusely, rather than concentrated at column densities favorable for shielding, molecular-gas formation, and star formation \citep{2009ApJ...693..216K, 2022ARA&A..60..319S}. We note that AGC\,983's ring-barred morphology (Section~\ref{sec:group}) may also contribute to its low azimuthally averaged surface density, independent of any feedback-driven redistribution. Thus, the suppressed star formation may reflect inefficient gas condensation and \HI-to-H$_2$ conversion rather than substantial removal of the atomic-gas reservoir. The central AGN could contribute by heating, stirring, or redistributing the cold gas through radiative or mechanical feedback \citep{2008ApJS..175..356H, 2012ARA&A..50..455F, 2024ApJ...971..165X}. The coexistence of a relatively normal total \HI\ mass and a low \HI\ surface density is therefore qualitatively consistent with feedback suppressing star formation before large-scale gas removal \citep{2025Natur.641.1137B}. However, the present observations do not establish a causal connection between the AGN and the diffuse \HI\ distribution, and other mechanisms that stabilize or redistribute the gas cannot be excluded.

Most galaxies in Figure~\ref{relation} (b) lie above the typical M$_*$-M$_{\textsc{Hi}}$ relation of \cite{2018ApJ...864...40P}, which is the more appropriate comparison given that our sample is itself largely selected via \HI\ detection (Section~\ref{sec:sample}). This ``gas-rich but star-formation-deficient" characteristic indicates a low global star formation efficiency relative to the atomic-gas reservoir. Since star formation is more directly associated with molecular gas than with the total \HI\ content \citep{2008AJ....136.2846B}, the spatial distribution and surface density of the atomic gas are important.


The resolved profiles in Figure~\ref{surface} show that, after accounting for the measurement uncertainties, the maximum azimuthally averaged \HI\ surface densities span approximately 5 to 20\,M$_{\odot}\,{\rm pc^{-2}}$. At the largest measured radii, the surface densities decline to a few M$_{\odot}\,{\rm pc^{-2}}$, approaching the galaxy-dependent 3$\sigma$ sensitivity limits of approximately 1--4\,M$_{\odot}\,{\rm pc^{-2}}$.
Nearby galaxies at sub-kpc resolution exhibit an empirical saturation of \HI\ at approximately 9\,M$_\odot\,{\rm pc^{-2}}$ \citep{2008AJ....136.2846B}, broadly consistent
with the characteristic \HI\ shielding column of $\sim$10~M$_\odot\,{\rm pc^{-2}}$ predicted for the atomic-to-molecular transition at solar metallicity \citep{2009ApJ...693..216K}. AGC\,1027, which has near-solar metallicity ([M/H]\,=−0.02; Table~\ref{tab:meerkat}), exceeds this characteristic surface density in its inner region, consistent with its high SFR (2.66\,M$_\odot$\,yr$^{-1}$, the highest in our sample).
The more metal-poor galaxies in our sample (e.g., AGC\,111629, AGC\,111930) likely require a higher threshold column for efficient H$_2$ formation, so their measured surface densities may be even further below the atomic-to-molecular transition. Their correspondingly low SFR ($\le$0.2\,M$_\odot$\,yr$^{-1}$) and low \HI\ surface density are consistent with genuinely low \HI-to-H$_2$ conversion in these systems, despite the lower spatial resolution ($\sim$4\,kpc) of our measurements.

AGC\,110319 is the only galaxy in the sample showing an apparent central decrease in its azimuthally averaged \HI\ surface-density profile. Its azimuthally averaged surface density rises from approximately 2.4~M$_\odot {\rm pc^{-2}}$ in the central annulus to a maximum of approximately 4.8~M$_\odot {\rm pc^{-2}}$ at a radius of $\sim$6 kpc, although this difference is not statistically significant and the depression is therefore tentative (Figure~\ref{surface}). This distribution differs from those reported for some other gas-rich UDGs, such as AGC\,114905 \citep{2024A&A...689A.344M} and KK\,176 \citep{2024A&A...689L...3L}, which retain appreciable central \HI\ reservoirs. Most gas-rich UDGs in the Finding Rotation In ultra-Diffuse gAlaxies (FRIDA) sample have shallow inner \HI\ profiles with central surface densities around 4\,M$_\odot\,{\rm pc}^{-2}$ and
an approximately exponential decline at larger radii \citep{2026arXiv261114635M}. 
AGC\,748816 also shows an overall outward decline, but from a higher central surface density of
7.8\,M$_\odot\,{\rm pc}^{-2}$, while AGC\,110319 appears to peak in
the second annulus, although its two innermost measurements agree
within their uncertainties (Figure~\ref{surface}).
The origin of the apparent central depression in AGC\,110319 remains uncertain. Its relatively isolated environment makes a strong ongoing tidal interaction less likely, but the present spatial resolution does not allow us to distinguish among feedback-driven gas redistribution, phase conversion, ionization, or other internal processes.

If confirmed, one possible explanation for the central \HI\ depression is gas redistribution driven by past stellar feedback, a mechanism proposed for kpc-scale \HI\ holes in nearby dwarf galaxies,  and for extended stellar and gas distributions in isolated UDGs \citep{2017MNRAS.466L...1D}. As noted in Section~\ref{sec:foreground}, the dust-corrected H$\alpha$-based SFR of AGC\,110319 is lower than the NUV-based value (0.034 versus 0.087~M$_\odot\,{\rm yr^{-1}}$). This is consistent with a recently declining star formation history, though the modest factor of $\sim$2.6 difference between the two SFR indicators does not require a strong past starburst. Given the limitation of current data, however, stellar feedback remains a plausible but unconfirmed interpretation. Alternatively, AGC\,110319 may illustrate diversity in the internal evolution of
gas-rich UDGs.

\subsection{Baryonic Tully-Fisher Relation}
\label{sec:btfr}

We show the BTFR in Figure~\ref{relation}(c). The baryonic mass is calculated as
 M$_{\rm bary}$=M$_*$+1.33M$_{\rm HI}$, where the factor of 1.33 accounts for helium \citep{2019MNRAS.483.1754D,2020MNRAS.495.3636M}. For the seven galaxies with usable \textsc{3DBarolo} models, we adopt the circular velocities or rotation velocities derived from the resolved \HI\ kinematics and propagate their final asymmetric uncertainties.

AGC\,110319 is excluded from the BTFR. Its fitted rotation speeds are comparable to or smaller than the gas dispersion, and the projected speeds fall below the channel spacing, so neither the tilted-ring fit nor the global line width provides a reliable rotation speed (Section~\ref{sec:modle}). Its position on the BTFR therefore cannot be constrained with the present data.

Most of the remaining galaxies lie broadly within the region bracketed by the reference relations for edge-on and low-mass gas-rich galaxies, once the velocity uncertainties are considered. The UDG AGC\,748816 shows the clearest upward displacement, with $V_{\rm rot}=68^{+20}_{-25}$\,\kms\ and $\log(M_{\rm bary}/M_\odot)\simeq9.60$. 
Its inclination, $i_{\rm HI}=29^\circ$, lies near the lower limit of the mock-observation tests that support our adopted inclination uncertainty (Section~\ref{sec:geometry}). Its inferred velocity is therefore particularly sensitive to uncertainties in disk geometry.
The \textsc{3DBarolo} fit of AGC\,748816 is also among the least well constrained in our sample, owing to diffuse and low-S/N position--velocity structure (Section~\ref{sec:modle}). Both factors should be borne in mind when interpreting its position on the BTFR.

AGC\,978 also lies moderately above the reference relations, with $V_{\rm circ}=103^{+15}_{-12}$\,\kms\ and $\log(M_{\rm bary}/M_\odot)\simeq10.31$. However, its displacement is smaller and becomes consistent with the edge-on relation toward the upper end of its velocity uncertainty. Importantly, AGC\,978 is not a UDG, indicating that an apparent upward BTFR offset is not unique to diffuse galaxies in this sample. If gas pressure declines outward, the uncorrected outer-ring rotation speeds of AGC\,111629, AGC\,748816, and AGC\,111930 provide lower limits on $V_{\rm circ}$, and accounting for pressure support would move AGC\,748816 toward the reference relations at fixed baryonic mass in Figure~\ref{relation}(c).

\section{Conclusion}
\label{sec:conc}

We have presented a detailed analysis of eleven galaxies in a group at redshift 0.02, along with two foreground galaxies at $z\sim0.016$, observed with MeerKAT as part of the MALS survey (pointing J0126+1420). 
The MeerKAT observations, with a synthesized beam of 12.1$''$$\times$7.2$''$ and channel width of 5.5\,\kms, provide spatially resolved \HI\ maps for eight galaxies.
By combining these data with multi-wavelength measurements, we examined their \HI\ morphologies, gas content, star formation activity, disk geometry, resolved kinematics, radial \HI\ surface-density profiles, and positions on the BTFR. Our main results are as follows.

\begin{enumerate}
\item The resolved \HI\ morphologies reveal a diverse galaxy population. Several systems show asymmetric, patchy, or extended \HI\ emission that may trace past or ongoing interactions, features not distinguishable in single-dish data. In contrast, the group UDG AGC\,748816 has a comparatively regular and approximately circular \HI\ distribution without clear evidence of a strong tidal disturbance. These differences demonstrate that galaxies in the same general environment can exhibit substantially different gas morphologies and evolutionary states.

\item Most of the \HI-detected galaxies lie above the \HI-selected M$_*$-M$_{\rm HI}$ scaling relation while having SFRs near or below the star-forming main sequence. They are therefore rich in atomic gas but have relatively low global star formation efficiencies. Accounting for their uncertainties, the maximum azimuthally averaged \HI\ surface densities span approximately 5--20~M$_{\odot}\,{\rm pc^{-2}}$, whereas the outermost detected regions generally reach only $\Sigma_{\rm HI}\lesssim$1--4~M$_{\odot}\,{\rm pc^{-2}}$. AGC\,1027, which has near-solar metallicity, exceeds the characteristic surface density associated with the atomic-to-molecular transition in its inner region, consistent with its high SFR. The more metal-poor galaxies in the sample remain below this scale, consistent with their low SFRs and reduced \HI-to-H$_2$ conversion efficiency.

\item AGC\,110319 is the only galaxy in the sample whose beam-matched profile
shows an apparent central decrease in \HI\ surface density, rising from
approximately 2.4~$\rm M_\odot\,pc^{-2}$ in the central annulus to
approximately 4.8~$\rm M_\odot\,pc^{-2}$ at a deprojected radius of
approximately 6~kpc. However, this difference is not statistically significant, and the 4.1$\times$2.4 kpc physical resolution provides only a limited number of independent resolution elements across this structure. The apparent central depression should therefore be regarded as tentative. The lower H$\alpha$-based SFR relative to the NUV-based value is consistent with recently declining star formation, but it does not demonstrate that feedback produced the apparent depression. Feedback-driven gas redistribution remains one possible explanation, alongside internal dynamics, ionization, and phase conversion. Higher-resolution \HI\ and spatially resolved multi-wavelength observations are required to distinguish among these possibilities.

\item Usable \textsc{3DBarolo} tilted-ring models are obtained for seven galaxies. Most of the measured velocities lie broadly within the region bracketed by the reference BTFRs when their uncertainties are considered. The group UDG AGC\,748816 shows the clearest upward displacement, with V$_{\rm rot}$=68$^{+20}_{-25}$~km\,s$^{-1}$, but its relatively low inclination, the limited quality of its resolved kinematic fit, and residual non-circular motions introduce important systematic uncertainties. AGC\,978, which is not a UDG, also shows a smaller upward displacement, indicating that such offsets are not unique to diffuse galaxies in this sample. The present measurements therefore do not provide compelling evidence for an anomalously low dark-matter content in AGC\,748816.

\end{enumerate}

These results demonstrate the importance of combining resolved \HI morphology, surface-density profiles, and three-dimensional kinematic modeling when interpreting gas-rich galaxies and UDGs. Deeper and higher-resolution \HI\ observations will be required to constrain non-circular motions and pressure support, while spatially resolved CO or other molecular-gas tracers are needed to determine how the extended atomic reservoirs are connected to star formation.

\section*{Acknowledgments}

We thank the anonymous referee for a detailed and constructive report, which substantially improved the analysis and presentation of this paper.
MeerKAT is operated by the South African Radio Astronomy Observatory, which is a facility of the National Research Foundation, an agency of the Department of Science and Innovation. 
The MeerKAT data were processed using the MALS computing facility at IUCAA (https://mals.iucaa.in/releases).
The Common Astronomy Software Applications (CASA) package is developed by an international consortium of scientists based at the National Radio Astronomical Observatory (NRAO), the European Southern Observatory (ESO), the National Astronomical Observatory of Japan (NAOJ), the Academia Sinica Institute of Astronomy and Astrophysics (ASIAA), the CSIRO division for Astronomy and Space Science (CASS), and the Netherlands Institute for Radio Astronomy (ASTRON) under the guidance of NRAO.
The National Radio Astronomy Observatory is a facility of the National Science Foundation operated under cooperative agreement by Associated Universities, Inc.
The NASA–Sloan Atlas is based on data from the Sloan Digital Sky Survey.
This work is supported by the National SKA Program of China (No. 2025SKA0150101), the National Natural Science Foundation of China, Grant nos. 12225303 and 12421003, the Chinese Academy of Sciences Project for Young Scientists in Basic Research, grant no. YSBR-063, and the New Cornerstone Science Foundation through the Xplorer Prize, the Guizhou Province High-level Talent Program (No.QKHPTRC-GCC[2022]003-2). 
P.C. acknowledges support from the Natural Science Foundation of Sichuan Province (No. 2026NSFSC0807) and the Specialized Research Fund for State Key Laboratory of Radio Astronomy and Technology (No. SDQZ-KF-A2613).
LEPM acknowledges the National Key R\&D Program of China (No. XY-2025-1459), the Science Fund for Creative Research Groups of the National Natural Science Foundation of China (No. 12221003), and the National Natural Science Foundation of China under grant No. 12573010.
%

\facilities{Arecibo, the Dark Energy Camera Legacy Survey (DECaLS), GALEX, MeerKAT and the Sloan Digital Sky Survey (SDSS).}

\software{
ARTIP \citep{Gupta21},
Astropy \citep{Astropy22},
CASA \citep{Casa22},
katbeam \citep{Mauch20},
Matplotlib \citep{Hunter07},
NumPy \citep{Harris20},
pPXF \citep{2004PASP..116..138C},
SciPy \citep{Virtanen2020SciPyPython},
SoFiA \citep[][]{Westmeier2021},
Cannubi \citep{2023MNRAS.521.1045R}, and
3DBarolo \citep{DiTeodoro2015}.
}

\appendix

\section{Additional data for the galaxies}
\label{app:additional_data}

Here, we provide MALS \HI\ source identifiers, SoFiA positions and morphological parameters, ALFALFA integrated fluxes, and local volume densities for the 13 sample galaxies, supplementing the main measurements presented in Table~\ref{tab:meerkat}.

\begin{sidewaystable*}
\caption{Additional \HI\ source properties and supporting measurements for the sample galaxies in the field of MALS J0126+1420.}
\centering
\footnotesize
\setlength{\tabcolsep}{3pt}
\begin{tabular}{c c c c c c c c c c c c}
\hline\hline
1 & 2 & 3 & 4 & 5 & 6 & 7 & 8 & 9 & 10\\
\hline
Source & MALS & RA & Dec & Major & Minor & PA & ALFALFA \HI\ flux$^{\dag}$ & $\rho_5$ & ($V_{\rm sys}$) \\
name &  ID & (J2000) & (J2000) & ($^{\prime\prime}$) & ($^{\prime\prime}$) & ($^\circ$) & (Jy\,km\,s$^{-1}$) & (Mpc$^{-3}$) & (km\,s$^{-1}$) \\
\hline
AGC\,112961$^{\dag}$ & J012507.20+143708.0 & 21.280000 & 14.618890 & -- & -- & -- & 0.71 $\pm$ 0.06 & 0.19 & --\\
AGC\,983 & J012521.91+145022.7 & 21.341300 $\pm$ 0.000321 & 14.839600 $\pm$ 0.000145 & 51.2 & 13.8 & $-$70.4 & 7.46 $\pm$ 0.12 & 0.24 & 6581 \\
IC\,0107$^{\dag}$ & J012524.72+145154.0 & 21.353000 & 14.865000 & -- & -- & -- & -- & 0.15 & -- \\
AGC\,978 & J012513.25+145221.9 & 21.305200 $\pm$ 0.000190 & 14.872800 $\pm$ 0.000216 & 18.8 & 14.9 & $-$24.7 & 4.24 $\pm$ 0.07 & 0.23 & 6368 \\
AGC\,1027 & J012709.49+144632.7 & 21.789500 $\pm$ 0.000129 & 14.775700 $\pm$ 0.000159 & 18.7 & 14.7 & 5.9 & 6.71 $\pm$ 0.11 & 0.44  & 6444\\
AGC\,110339$^{\dag}$ & J012730.89+144910.0 & 21.878700 & 14.819440 & -- & -- & -- & 0.90 $\pm$ 0.06 & 0.35 & --\\
AGC\,111629 & J012645.38+143728.5 & 21.689100 $\pm$ 0.000137 & 14.624600 $\pm$ 0.000216 & 18.9 & 9.9 & 15.8 & 1.84 $\pm$ 0.08 & 0.43 & 6453\\
AGC\,748815$^{\dag}$ & J012702.69+143938.0 & 21.761200 & 14.660550 & -- & -- & -- & 0.94 $\pm$ 0.07 & 0.39  & --\\
AGC\,748816 & J012756.78+143008.5 & 21.986600 $\pm$ 0.000199 & 14.502400 $\pm$ 0.000222 & 17.8 & 15.0 & $-$19.0 & 1.41 $\pm$ 0.06 & 0.41 & 6406\\
AGC\,111930 & J012454.59+141307.5 & 21.227400 $\pm$ 0.000175 & 14.218700 $\pm$ 0.000177 & 16.9 & 10.7 & $-$48.3 & 1.64 $\pm$ 0.09 & 0.10 & 5914\\
AGC\,112965$^{\ddag}$ & J012528.40+141148.4 & 21.368300 & 14.196800 & -- & -- & -- & 0.48 $\pm$ 0.05 & 0.11 & --\\
AGC\,111931 & J012504.28+142402.4 & 21.267800 $\pm$ 0.000270 & 14.400700 $\pm$ 0.000123 & 33.8 & 11.1 & $-$76.1 & 3.46 $\pm$ 0.08 & 0.028 & 5036\\
AGC\,110319 & J012516.62+140853.7 & 21.319300 $\pm$ 0.000264 & 14.148200 $\pm$ 0.000218 & 24.1 & 17.9 & $-$65.8 & 1.41 $\pm$ 0.06 & 0.031 & 5041\\
\hline
\end{tabular}
\normalsize
\begin{tablenotes}
\small
\item Columns 1 and 2: Source names from ALFALFA or the Index Catalog \citep{1895MmRAS..51..185D} and the corresponding MALS \HI\ identifier, in the same order as Table~\ref{tab:meerkat}. Targets marked $\dag$ are ALFALFA-detected but MeerKAT-undetected; targets marked $\ddag$ were detected by MeerKAT at insufficient S/N for morphological analysis. 
Columns 3 and 4: \HI\ flux-weighted RA and Dec, and their 1$\sigma$ positional uncertainties. For sources not detected in \HI, the MALS identifier, RA and Dec are based on the optical or AGC position.
Columns 5, 6 and 7: Major axis, minor axis, and position angle of the elliptical fit to the \HI\ morphology, derived using \textsc{SoFiA}, respectively. 
Column 8: Integrated \HI\ flux density from the ALFALFA$^{\dag}$ survey \citep{2018ApJ...861...49H}. IC\,0107 is undetected in the ALFALFA.
Column 9: Volume density to the 5th nearest neighbor (see Section~\ref{sec:density}).
Column 10: Heliocentric systemic velocity of each galaxy from the MeerKAT global \HI\ profile (see Section~\ref{sec:modle}).
\end{tablenotes}
\label{tab:additional_data}
\end{sidewaystable*}

\clearpage

\section{\HI\ inclination from beam-corrected isophotes}
\label{app:hi_geometry_method}

The surface-brightness level of each fitted isophote was defined relative
to the peak of the smoothed masked moment-0 map as
\begin{equation}
M_{\rm iso}
=
f M_{0,\rm peak},
\end{equation}
where \(0<f<1\) is the adopted fractional isophotal level.

To correct the fitted ellipse for the finite spatial resolution, we
represented it by an equivalent two-dimensional covariance matrix. For
an elliptical Gaussian,
\begin{equation}
\frac{I(\boldsymbol{x})}{I_0}
=
\exp\left[
-\frac{1}{2}
\boldsymbol{x}^{\mathsf{T}}
\mathbf{C}^{-1}
\boldsymbol{x}
\right]
\end{equation}
An isophote at a fractional intensity \(f=I/I_0\) therefore corresponds
to the dimensionless elliptical radius
\begin{equation}
\rho(f)
=
\sqrt{-2\ln f}.
\end{equation}
For a fitted ellipse with semi-major axis \(a\), semi-minor axis \(b\),
and position angle \(\phi\), the equivalent observed covariance matrix
is
\begin{equation}
\mathbf{C}_{\rm obs}
=
\mathbf{R}(\phi)
\left(
\begin{array}{cc}
\left(a/\rho\right)^2 & 0 \\[3pt]
0 & \left(b/\rho\right)^2
\end{array}
\right)
\mathbf{R}^{\mathsf{T}}(\phi),
\end{equation}
where \(\mathbf{R}(\phi)\) is the corresponding rotation matrix.

Under the Gaussian-convolution approximation, the intrinsic covariance
matrix was calculated as
\begin{equation}
\mathbf{C}_{\rm int}
=
\mathbf{C}_{\rm obs}
-
\mathbf{C}_{\rm beam}
-
\sigma_{\rm sm}^{2}\mathbf{I},
\end{equation}
where \(\mathbf{C}_{\rm beam}\) is the covariance matrix of the
elliptical synthesized beam, \(\sigma_{\rm sm}\) is the standard
deviation of the additional circular smoothing kernel, and
\(\mathbf{I}\) is the identity matrix. The beam major and minor axes were
converted from FWHM values to Gaussian standard deviations before
constructing \(\mathbf{C}_{\rm beam}\).

The intrinsic projected axis ratio was obtained from the positive
eigenvalues of \(\mathbf{C}_{\rm int}\),
\begin{equation}
q_{\rm HI}
=
\sqrt{
\frac{\lambda_{\rm min}}
     {\lambda_{\rm max}}
     },
\end{equation}
where \(\lambda_{\rm min}\) and \(\lambda_{\rm max}\) are the smaller
and larger eigenvalues, respectively. Realizations with non-positive
eigenvalues were rejected because they correspond to an unresolved or
unphysical beam-deconvolved ellipse.

Finally, the projected axis ratio was converted into an inclination by
assuming an oblate disk with an intrinsic thickness \(q_0=0.2\),
\begin{equation}
\cos^{2}i_{\rm HI}
=
\frac{
q_{\rm HI}^{2}-q_0^{2}
}{
1-q_0^{2}
}.
\end{equation}

\section{Deprojected galactocentric radius and H\,{\sc i} surface density}
\label{app:hi_profiles}

For each pixel, we calculated the eastward and northward angular
offsets, $\Delta E$ and $\Delta N$, from the adopted galaxy centre.
The offsets were rotated into directions parallel and perpendicular
to the disk major axis:

\begin{equation}
X = \Delta E\sin({\rm PA})+\Delta N\cos({\rm PA}),
\end{equation}

\begin{equation}
Y = \Delta E\cos({\rm PA})-\Delta N\sin({\rm PA}),
\end{equation}

where PA is measured from north through east. Under the assumption
that the H\,{\sc i} disk is intrinsically circular, the deprojected
angular radius is

\begin{equation}
R_{\rm arcsec}
=
\left[
X^{2}
+
\left(\frac{Y}{\cos i}\right)^{2}
\right]^{1/2}.
\label{eq:deprojected_radius_radius}
\end{equation}

This is equivalent to using elliptical annuli with a projected axis
ratio $q=\cos i$ and follows the standard tilted-ring formalism
\citep{Begeman1989,DiTeodoro2015,Oh2018}.



The corresponding line-of-sight H\,{\sc i} surface density is

\begin{equation}
\Sigma_{\rm HI,los}
=
\frac{N_{\rm HI}}{1.249\times10^{20}}
\quad {\rm M_{\odot}\,pc^{-2}}.
\end{equation}

We corrected this value to a face-on surface density using \citep{Leroy2008} 

\begin{equation}
\Sigma_{\rm HI}
=
\Sigma_{\rm HI,los}\cos i.
\label{eq:faceon_sigma}
\end{equation}

\bibliographystyle{apj}
\bibstyle{thesisstyle}
\bibliography{main}

\end{document}